\documentclass[prd,aps,preprint]{revtex4}
\usepackage{graphicx}
\usepackage{dcolumn}
\usepackage{amsfonts}

\def \nn{\nonumber}
\newcommand{\na}{\nabla}
\newcommand{\ep}{\epsilon}

\newcommand{\vphi}{\varphi}

\newcommand{\pa}{\partial}
\newcommand{\td}{\tilde}
\newcommand{\sla}[1]{\slash\!\!\! #1}

\newcommand{\beq}[1]{\begin{eqnarray}\label{#1}}
\newcommand{\eeq}{\end{eqnarray}}

\newcommand{\CN}[1]{{\cal N}=#1}

\begin{document}
\title{BPS $R$-balls in $\CN{4}$ SYM on
$R \times S^3$, \\
Quantum Hall Analogy and AdS/CFT Holography}
\author{Jian Dai$^{1,2}$}\email{jdai@physics.utah.edu}
\author{Xiao-Jun Wang$^{1,3}$}\email{wangxj@ustc.edu.cn}
\author{Yong-Shi Wu$^1$}\email{wu@physics.utah.edu}
\affiliation{\centerline{$^1$ Department of Physics, University of
Utah, Salt Lake City, Utah 84112, USA} \centerline{$^2$ Physics
Department, City College of CUNY, 160 Convent Ave., New York, NY
10031, USA} \centerline{$^3$ Interdisciplinary Center for
Theoretical Study} \centerline{University of Science and
Technology of China, AnHui, HeFei 230026, China} }
\begin{abstract}
In this paper, we propose a new approach to study the BPS dynamics
in $\CN{4}$ supersymmetric $U(N)$ Yang-Mills theory on $R \times
S^3$, in order to better understand the emergence of gravity in
the gauge theory. Our approach is based on supersymmetric,
space-filling $Q$-balls with $R$-charge, which we call $R$-balls.
The usual collective coordinate method for non-topological scalar
solitons is applied to quantize the half and quarter BPS
$R$-balls. In each case, a different quantization method is also
applied to confirm the results from the collective coordinate
quantization. For finite $N$, the half BPS $R$-balls with a $U(1)$
$R$-charge have a moduli space which, upon quantization, results
in the states of a quantum Hall droplet with filling factor
$\nu=1$. These states are known to correspond to the ``sources''
in the Lin-Lunin-Maldacena geometries in IIB supergravity. For
large $N$, we find a new class of quarter BPS $R$-balls with a
non-commutativity parameter. Quantization on the moduli space of
such $R$-balls gives rise to a non-commutative Chern-Simons matrix
mechanics, which is known to describe a fractional quantum Hall
system. In view of AdS/CFT holography, this demonstrates a
profound connection of emergent quantum gravity with
non-commutative geometry, of which the quantum Hall effect is a
special case.
\end{abstract}
\preprint{USTC-ICTS-06-02} \maketitle

\section{Introduction}

The AdS/CFT correspondence \cite{AdS-CFT} is a gravity/gauge
theory duality, according to which gravity is an emergent
phenomenon in the dual gauge field theory. In the most
well-understood case, the classical $AdS_5\times S^5$ geometry is
conjectured to be encoded in the strong 't Hooft coupling regime
of the $\CN{4}$ supersymmetric Yang-Mills (SYM) theory on $R
\times S^3$ with gauge group $U(N)$. A thorough understanding of
how this correspondence happens remains a challenge.

In the last two years, there has been encouraging progress. In
ref.~\cite{Ber04a}, Berenstein first proposed to consider a
decoupled limit which singles out the half BPS sector in the
$\CN{4}$ SYM. As a model for the dynamics of the half BPS states,
he studied the gauged mechanics of a holomorphic normal matrix,
which was shown to be equivalent to a one dimensional free fermion
system in phase space. (Similar results had been obtained in a
complex matrix model~\cite{CJR}.) The 1-d free fermion system can
be mapped to an integer quantum Hall (IQH) droplet in two
dimensions. An amazing and profound connection of the IQH droplet
picture with type IIB geometries on the gravity side was
subsequently revealed in a seminar paper by Lin, Lunin and
Maldacena (LLM) \cite{LLM}. By solving the equations of motion in
IIB supergravity in ten dimensions, they have been able to obtain
all non-singular half BPS type IIB geometries with isometry
$R\times SO(4)\times SO(4)$, which turn out to be completely
determined by the boundary value of a single real function $z$ on
a plane. The boundary value of $z$ can be only $\pm 1/2$, which
may be interpreted as the distributions of two types of point
charge sources on the boundary plane. Either region with $z=1/2$
or $z=-1/2$ can be viewed as a droplet of an incompressible fluid,
like the IQH fluid. The simplest case is the familiar geometry
$AdS_5\times S^5$, which corresponds to a circular droplet on the
boundary plane. In this way, one is tempted to associate the half
BPS geometries in IIB supergravity with the half BPS states in
$\CN{4}$ SYM, by comparing the boundary IQH droplets in LLM's half
BPS geometry to those in the phase space of Berenstein fermions.
This comparison is justified only when one can make sense to
Berenstein fermions in the half BPS sectors in $\CN{4}$ SYM, not
merely in a plausible model. Attempts to substantiate this
comparison have been made in several recent papers
\cite{MR05,TT05} with various degrees of success. (For the recent
generalization of gauged matrix mechanics to 1/4 and 1/8 BPS
states in $\CN{4}$ SYM, see \cite{Ber05a}.)

In this paper we will directly attack the problem of how to see
the Berenstein fermions emerging in the $\CN{4}$ SYM by examining
the BPS scalar field backgrounds. Our motivation came from the
recognition that the emergence of gravity in SYM, as the
correspondence between Berenstein and LLM fermions should imply,
actually indicates the background independence on both sides of
the holographic duality. To test the background independence, it
is certainly desirable to have a systematic approach for finding
candidate states on the gauge theory side for possible (classical
or quantum) geometries on the gravity side. The simplest candidate
states in SYM are the quantum BPS states with sufficiently large
$R$-charge. For their explicit construction, we propose a new
approach, which starts with constructing classical BPS backgrounds
in SYM, namely solutions to the equations of motion that saturate
the BPS bound and maintain a fraction of supersymmetry. The BPS
properties are required again because the AdS/CFT duality is a
strong-weak coupling duality; to check it one needs to be able to
interpolate between the weak and strong couplings, and a fraction
of unbroken supersymmetry might allow one to do so. The classical
BPS backgrounds with given conserved charges usually have a moduli
space for their collective coordinates. We suggest to do
quantization on the moduli space of collective coordinates and
expect that at least part of the resulting quantum states are BPS
protected candidate states that we are looking for.

We restrict ourselves in this paper to scalar backgrounds, which
can be viewed as Bose-Einstein condensates (BEC). It has been an
old folklore in theoretical physics community that classical
geometry, in a certain sense, can be viewed as a sort of BEC. So
we think it natural to relate the geometric backgrounds in gravity
to the scalar backgrounds in the SYM dual. For $\CN{4}$ SYM on $R
\times S^3$, the classical scalar vacuum is unique because of the
conformal coupling to the curvature of $S^3$, so the only
available BEC-like objects are space-filling $Q$-balls
\cite{Coleman85} with conserved $R$-charge; we call them simply
$R$-balls. The above considerations led us to examine the $R$-ball
solutions that preserve $1/2$, $1/4$ or even less supersymmetry.
To formulate the $R$-ball approach in this paper, we aim at 1)
formulating the general conditions for BPS backgrounds in SYM; 2)
finding explicitly classical BPS $R$-balls, particularly new
solutions corresponding to non-commutative geometry in the large
$N$ limit; 3) carrying out collective-coordinate quantization on
the moduli space of certain BPS $R$-balls; 4) showing the
emergence of the known IQH droplet of fermions in a certain sector
of half BPS $R$-ball configurations; and 5) finally showing that
the new non-commutative BPS $R$-balls, upon quantization, lead to
fractional quantum Hall (FQH) states, thus lending support to our
previous argument \cite{DWW05} for the possible appearances of
FQH-like states on the gravity side.

The success in carrying out the above steps 1)-3) will lay down
the foundation for a general framework for finding classical BPS
backgrounds in $\CN{4}$ SYM on $R\times S^3$, and for constructing
the quantum BPS states living on their moduli space (with given
conserved charges), particularly in the large $N$ limit. It is
expected that the distinctive features due to the fact that the
conformal field theory is defined on a compact space $S^3$ will
play a decisive role in these discussions. The above points 4) and
5) will dwell on the relevance of the IQH and FQH states in the
context of the holographic gravity/gauge duality. It has been
noticed in the literature that the many-fermion state in phase
space looks like an IQH droplet. Despite this, there is a crucial
difference between the two, since the IQH droplet pre-requires the
presence of the Landau levels and the projection to the lowest
Landau level, while the former does not. If indeed it makes sense
to talk about ``Landau levels'', then new states other than the
IQH droplet can emerge due to the inclusion of interactions,
giving rise to new candidates for non-perturbative states on the
gravity side. The authors of the present paper have put forward
arguments from the gravity side supporting the emergence of
FQH-like states \cite{DWW05}. The essence of the arguments was the
following: The interactions between the giant graviton probes in
the LLM geometry background are shown to be repulsive; if the
interactions can be extrapolated to finite density, then the giant
gravitons in the LLM geometry at right densities can condense into
new incompressible QH fluids with fractional filling factors. More
concretely, the dynamics of giant graviton probes is first shown
\cite{DWW05} to be described by a non-commutative Chern-Simons
gauge theory \cite{Susskind01a}. Then it was further reduced to a
non-commutative Chern-Simons matrix mechanics (NCCSMM) previously
proposed in ref. \cite{Poly01}, and its spectrum was shown to
contain not only the IQH but also the FQH states. This has
inspired us to try to find the FQH-like states on the gauge theory
side. But this did not seem easy in the matrix model approach
\cite{Ber04a,Ber04b}. (For other effort in studying the QHE in
SYM, see refs. \cite{GMSS05,Yamada05}.) Actually this was our main
motivation to look for a new approach to the BPS dynamics in
$\CN{4}$ SYM. Indeed as shown below, new non-commutative BPS
backgrounds can be found in our new framework in the large $N$
limit, confirming the relevance of the FQHE and, more generally,
of the non-commutative geometry to emergent gravity in SYM.

This paper is organized as follows. We start with a brief review
of supersymmetry transformations in $\CN{4}$ SYM on $R \times S^3$
in Sec. II, to set up the notations and to formulate the
conditions for unbroken supersymmetry. In Sec. III, we present
some variational theorems for the classical BPS $R$-balls, i.e.
classical scalar configurations of the lowest energy in the sector
with given $R$-charges and leaving part of the supersymmetry
unbroken. Then in Sec. IV, we proceed to construct the explicit
solutions for classical $R$-balls, which include the commutative
half BPS configurations known in the literature. In particular, we
find that in the large $N$ limit there may exist non-commutative
solutions, which solve the Gauss's constraints exactly and satisfy
the BPS bound with an error of order $O(1/N)$. Thus the moduli
space of the BPS $R$-balls is enhanced in the large $N$ limit. We
demonstrate this by presenting a new family of quarter BPS
$R$-balls that involve a non-commutativity parameter between
scalar and pseudo-scalar in pairs. In Sec. V, we discuss first in
great detail the collective coordinate (or moduli space)
quantization of the commutative half BPS $R$-balls. In particular,
besides demonstrating how our new approach reproduces the known
results in the half BSP sector, we show that the quantization
naturally leads to the ``Landau-level problem'', so that it makes
sense to look for more exotic FQH (or FQH-like) states in more
complicated sectors. Indeed in Sec. VI we are able to show that
the quantization of the non-commutative quarter BPS $R$-balls
leads to an NCCSMM model for infinite-dimensional matrices, whose
Hilbert space indeed contains FQH-like states. Sec. VII is devoted
to summary and discussions. Finally, in Appendix of this paper, we
derive the $\CN{4}$ supersymmetry algebra on $R \times S^3$ and
present the formula for the BPS bound, which assures that for our
$R$-balls, the BPS bound is saturated by their $R$-charge, exactly
corresponding to the BPS bound in the gravity dual.

\section{$\CN{4}$ SYM on $R \times S^3$}

It is known that the $\CN{4}$ supersymmetric vector multiplet in
four dimensions can couple to a background metric in a Weyl
invariant manner classically \cite{BF82}. However, in quantum
theory there exists Weyl anomaly except for certain symmetric
backgrounds, such as $R \times S^3$ \cite{Okuyama02}. Accordingly,
the $\CN{4}$ SYM on $R \times S^3$ is a well-defined quantum
conformal field theory; the corresponding Lagrangian in $\CN{1}$
language reads
\beq{1} {\cal L}&=&-\frac{1}{4}{\rm Tr}(F_{\mu\nu}F^{\mu\nu})
+\frac{1}{2} {\rm
Tr}(D_\mu X_iD^\mu X^i +D_\mu Y_iD^\mu Y^i)-V(X,Y) \nn \\
&& +\frac{i}{2}{\rm Tr} (\bar{\psi}\gamma^a e^\mu_a
D_\mu\psi)-\frac{ig}{2}{\rm Tr}\{\bar{\psi}(\alpha^i[X_i,\psi]
+\gamma^5\beta^j[Y_j,\psi])\}. \eeq Here $g$ is the gauge
coupling; the potential of scalars is given by
\beq{2}V(X,Y)=\frac{1}{2R^2}{\rm Tr}(X_iX^i+Y_jY^j)
-\frac{g^2}{4}{\rm Tr}([X_i,X_j]^2+[Y_i,Y_j]^2+2[X_i,Y_j]^2) \eeq
with $R$ the radius of $S^3$; $X$ and $Y$ denote scalar and
pseudo-scalar fields respectively.
$\alpha^i,\;\beta^j\;(i,j=1,2,3)$ are $4\times 4$ real
anti-symmetric matrices satisfying the algebraic relations
\cite{Osborn79} \beq{3}
[\alpha^i,\alpha^j]&=&-2\ep^{ijk}\alpha^k,\hspace{0.5in}
[\beta^i,\beta^j]=-2\ep^{ijk}\beta^k, \nn \\
\{\alpha^i,\alpha^j\}&=&\{\beta^i,\beta^j\}=-2\delta^{ij},
\hspace{0.5in}[\alpha^i,\beta^j]=0. \eeq All gamma matrices are
defined in a local Lorentz frame $e_a^\mu$ and $D_\mu$ is the
covariant derivative: \beq{4} D_\mu\psi=\na_\mu\psi-ig[A_\mu,\psi]
=\pa_\mu\psi+\Omega_\mu\psi-ig[A_\mu,\psi], \eeq where
$\Omega_\mu=\frac{1}{4}\Omega^{ab}_\mu\gamma_{ab}$ is the spin
connection with $\Omega^{ab}$ the connection 1-form with
$\gamma_{ab}$ the generators of local Lorentz transformations. Two
remarks are deserved here. First, it is the cylinder $R \times
S^3$, instead of Minkowski space $M^4$, that is the global
conformal boundary of $AdS_5$. Second, because of the cylindrical
structure of $R \times S^3$, the natural spin connection does not
involve the temporal direction, implying $\nabla_0=\partial_0$; as
a result, only local $SO(3)$ transformations, instead of local
$SO(3,1)$ transformations, act upon the gaugino field $\psi$;
namely, there are no local boosts. For this reason, a global and
Majorana formalism for $\psi$ and $\gamma_a$ can be defined on $R
\times S^3$ similar to that in flat space. More subtleties of the
spin structure on $R \times S^3$ will be explored in the following
discussion of global supersymmetry.

The action integral of the Lagrangian~(\ref{1}) over $R \times
S^3$ possesses an $\CN{4}$ superconformal symmetry because of the
existence of four conformal Killing spinors $\ep_A$ for
$A=1,2,3,4$. We will first follow the analysis in refs.
\cite{BDPS87,NST}, which dealt with the conformal Killing spinors
on $R \times S^3$ by descending from the Killing spinors on
$AdS_5$. In this formalism, the conformal Killing spinor equations
are written as \beq{5}
\pa_0\ep_A=\frac{i}{2R}\Gamma_0\ep_A,\hspace{0.5in}
\na_m\ep_A=\frac{i}{2R}\Gamma_m\Gamma_5\ep_A \ , \eeq where
$m=1,2,3$ labels the directions in a local orthonormal frame on
$S^3$, and the five upper-cased gamma matrices generate the
Clifford algebra of five-dimensional Minkowski space $M^5$, which
form a local frame on $AdS_5$. To the best, there are four
linear-independent complex global sections in the spin bundle on
$R \times S^3$ as the solutions to Eq.~(\ref{5}) for each
$A=1,2,3,4$; therefore, there are at most $32$ supercharges.

In addition to the above {\em extrinsic} formalism (descending
from $AdS_5$ in Eq.~(\ref{5})), there is an {\em intrinsic}
formalism for the spin structure on $R \times S^3$, by making use
of Majorana spinors, in which the gamma matrices are denoted with
the lower case. In fact, the two formalisms share the following
feature: There is only one local $SO(3)$ acting on the
four-component complex spinor $\ep_A$. This implies that the
four-component spinor $\ep_A$ in Eq.~(\ref{5}) as a representation
of this local symmetry must be reducible. A natural reduction
results from the observation that $[\Gamma_0,\Gamma_m\Gamma_5]=0$.
One can introduce a projection operator
$\mathcal{P}=(1+i\Gamma_0)/2$ such that a ``$\Gamma_0$-chirality''
is defined as follows: (for convenience, the subscript $A$ is
omitted for a moment)
\begin{equation}
 \ep_L:=\mathcal{P}\ep,~~\ep_R:=(1-\mathcal{P})\ep.
\end{equation}
Two Majorana spinors in the intrinsic spin structure can be
constructed from $\ep_L$ or $\ep_R$, respectively, via the
standard procedure that produces a Majorana spinor from a Weyl
spinor:
\beq{6} \zeta_L=\ep_L+\mathcal{C}\ep_L^\ast, ~~
\zeta_R=\ep_L+\mathcal{C}\ep_R^\ast , \eeq where $\mathcal{C}$ is
the conventional charge conjugation matrix. Note that the
conformal Killing spinor $\ep$ in the (extrinsic) $AdS$-descending
spin structure does not admit any Majorana condition.

With the above-mentioned clarification of the spin structure on $R
\times S^3$, we can write down the fermionic part of $\CN{4}$
superconformal transformation explicitly: \cite{NST,Okuyama02}:
\beq{susy} \delta_{L,R}
A_\mu&=&-i\bar{\psi}\gamma_\mu\zeta_{L,R},\hspace{0.5in}
\delta_{L,R} X_i=\bar{\psi}\alpha^i\zeta_{L,R}, \hspace{0.5in}
\delta_{L,R} Y_j=i\bar{\psi}\beta^j\gamma_5\zeta_{L,R}, \nn \\
\delta_{L,R}\psi&=&\frac{1}{2}\gamma^{\mu\nu}F_{\mu\nu}\zeta_{L,R}
-i\gamma^\mu(\alpha^iD_\mu X_i+i\gamma_5\beta^jD_\mu
Y_j)\zeta_{L,R}
\nn \\
&&+\frac{ig}{2}\ep^{ijk}(\alpha^k[X_i,X_j]+\beta^k[Y_i,Y_j])
\zeta_{L,R}
+g[X_i,Y_j]\alpha^i\beta^j\gamma_5\zeta_{L,R}, \nn \\
&&-\frac{i}{2}(\alpha^iX_i-i\gamma_5\beta^jY_j)
\gamma^\mu\na_\mu\zeta_{L,R} \ , \eeq where, as usual, $\gamma^\mu
=\gamma^a e_a^\mu$ with $e_a^\mu$ the local vierbein. As $R$ is
sent to infinity, both the Lagrangian~(\ref{1}) and the
superconformal transformation~(\ref{susy}) reduce to those in
$M^4$. Using the conformal Killing spinor equations (\ref{5}), we
have \beq{8} \gamma^\mu\na_\mu\zeta_L=
\frac{2i}{R}\gamma_5\gamma^0\zeta_L,\hspace{0.6in}
\gamma^\mu\na_\mu\zeta_R=-\frac{2i}{R}\gamma_5\gamma^0\zeta_R .
\eeq Then one can directly check that the variation of the
Lagrangian~(\ref{1}) is indeed a total derivative. We will show in
Appendix that either $\zeta_L$ or $\zeta_R$ generates a
super-isometry algebra separately; together, they generate the
entire superconformal algebra.

\section{$R$-Balls as Classical BPS Backgrounds}

As mentioned in the introduction, recent progress in understanding
emergent gravity in AdS/SYM holography motivated us to look for
scalar field configurations (representing a sort of BEC) that
preserve a fraction of supersymmetry. In this attempt, the old
idea of Coleman's $Q$-balls \cite{Coleman85} has attracted our
attention. Initially, $Q$-balls are defined as non-topological
soliton solutions in complex scalar field theories in a
four-dimensional flat spacetime. Their existence and classical
stability hinge on the existence of a conserved charge, $Q$,
associated with a global $U(1)$ symmetry: A $Q$-ball is the
solution minimizing the energy in the sector with a fixed and
sufficiently large $Q$-charge. The solutions constructed in ref.
\cite{Coleman85} are spherically symmetric in space, and the
nonzero $Q$-charge is generated by rotating a static configuration
in internal space. In our present case, we generalize Q-ball to
the compact space $S^3$; this generalization allows the existence
of {\em space-filling} $Q$-balls, which is impossible in
non-compact flat space. In $\CN{4}$ SYM, there are two important
complications for $Q$-balls. First, the pertinent global symmetry
is $SO(6)$ $R$-symmetry. The corresponding $Q$-balls, which we
call {\em $R$-balls}, carry a $U(1)$ $R$-charge embedded in the
non-abelian $SO(6)$. Different embedding yields a different type
of $R$-balls. Second, the theory has a color $U(N)$ gauge
symmetry, and the scalar fields also carry color degrees of
freedom. The Gauss's law will severely constrain possible physical
states after quantization. In this section, we will formulate and
analyze the conditions for the classical $R$-balls, namely
solutions to the equations of motion with energy saturated by the
$R$-charge in a given sector with fixed $R$-charge.

\subsection{Variational BPS Bound}

First we consider the variational aspects of the problem. As usual
for classical backgrounds, we set the gaugino field to zero:
$\psi=0$. Then the Hamiltonian reads:
\beq{hamiltonian}
H=\int\limits_{S^3}\Bigl({1\over 2}{\rm Tr} (F_{0i}^2+ {1\over
2}F_{ij}^2+D_i\phi_sD_i\phi_s +D_0\phi_sD_0\phi_s +{1\over
R^2}\phi_s\phi_s) +V_4(\phi)\Bigr),
\eeq
where $\{\phi_s\}=\{X_a,Y_a\}$ (with $s=1, \cdots, 6$ and
$a=1,2,3$) are six scalar fields, transforming as a vector under
$R$-symmetry $SO(6)$, while as adjoint representation under color
$U(N)$. We note that all terms in Eq.~(\ref{hamiltonian}) are
non-negative. To look for the $Q$-ball solutions, we concentrate
on the dynamics of the scalar fields. Thus we set $F_{\mu\nu}=0,
(\mu,\nu=0,1,2,3)$, putting the first two terms to zero. Because
$S^3$ is simply connected, it is possible to take the spatial
components of the gauge potential $A_i=0$, while allowing $A_0$ a
function of time only. In the following, we will adapt the BPS
analysis to the global $R$-symmetry.

Let us focus on a sector of scalar configurations with a fixed
$U(1)$ $R$-charge: \beq{R-charge} Q_\mathbf{r} = {\rm
Tr}\int_{S^3} (D_0\phi_s) \mathbf{r}_{st} \phi_t, \eeq associated
with a generator $\mathbf{r}$ in the $so(6)$ Lie algebra (in the
definition representation). Note that as a six-by-six {\em
antisymmetric} matrix, $\mathbf{r}=(\mathbf{r}_{st})$ may be
degenerate. If $\mathbf{r}$ is also an {\em orthogonal} matrix in
a linear subspace in which it is non-degenerate, then it is easy
to prove the following BPS-like inequality:
\begin{equation}\label{BE}
 H\ge {1\over R}\Bigl|{\rm Tr}\int_{S^3}
 (D_0\phi_s) \mathbf{r}_{st} \phi_t\Bigr|
={|Q_\mathbf{r}|\over R}.
\end{equation}
The energy $H$ saturates the lower bound set by the charge
$Q_{\mathbf{r}}$ in (\ref{BE}) only when the following three
conditions are satisfied: First, \beq{BPScDi} D_i\phi_s&=&0, \quad
(i=1,2,3), \eeq which makes the third term in
Eq.~(\ref{hamiltonian}) vanish; and
\begin{eqnarray}\label{BPScC}
 [\phi_s,\phi_t]&=&0,
\end{eqnarray}
making the contribution of the quartic potential $V_4$ vanish; and
finally
\begin{eqnarray}\label{BPScD0}
 D_0\phi_s&=&\pm R^{-1}\mathbf{r}_{st}\phi_t,
\end{eqnarray}
with $\phi_s$ having {\em only non-zero} components in the
subspace in which the generator $\mathbf{r}$ is non-degenerate.
Since $A_i=0$, (\ref{BPScDi}) implies that the scalar fields
$\phi_s$ are constant, i.e. the lowest KK modes, on $S^3$. The
condition (\ref{BPScD0}) means the time-dependent configuration
$\phi_s(t)$ rotates in internal space with a specific frequency,
which generates the $R$-charge that saturates the lower energy
bound. Note that after imposing $F_{\mu\nu}=0$ and
Eq.~(\ref{BPScDi}), the Hamiltonian is reduced to a gauged
mechanics: \beq{hamiltonian1} H= \int\limits_{S^3}
\Bigl\{D_0\phi_s D_0\phi_s +{1\over R^2}\phi_s \phi_s +V_4(\phi)
\Bigr\} . \eeq The conditions (\ref{BPScDi}) and (\ref{BPScC}), we
will call the BPS conditions, are those to saturate the energy
bound in a given sector with definite $R$-charge.

In the next section, we will see that the configurations
satisfying these BPS conditions automatically preserve part of the
$\CN{4}$ supersymmetry. Moreover, we note that when the
commutative condition (\ref{BPScC}) is satisfied, the Gauss's law
constraint is also automatically satisfied, since the color charge
density vanishes:
\begin{equation}
 j^0_{U(N)}=[D_0\phi_s,\phi_s]
=\pm R^{-1}\mathbf{r}_{st} [\phi_t,\phi_s]=0.
\end{equation}

We will call the $Q$-ball solutions to the equations of motion
obtained by solving the BPS equations (\ref{BPScD0}) and
({\ref{BPScC}) as BPS $R$-balls. As zero-modes on $S^3$, they are
space-filling. They form a decoupled sector in the limit when the
radius $R$ of $S^3$ tends to zero. (Note that non-space-filling
$R$-ball configurations exist on $S^3$, but they may not be BPS in
the sense that the energy is not saturated by their $R$-charge.
This is in accordance with ref. \cite{BBFH}. In the same limit,
the gaugino and gluon backgrounds are decoupled from this sector
too \cite{Okuyama02}.) Here we would like to warn that the BPS
Eq.~(\ref{BPScD0}) does not have a topological origin, since the
$R$-charge is not central in SUSY algebra. Incidentally, we also
make the remark that topological BPS solitons, such as 't
Hooft-Polyakov monopoles and dyons with non-vanishing charges, do
not exist on the compact $S^3$. This is the main reason why we
turn our attention to non-topological $R$-balls in search of
BEC-like backgrounds.

\subsection{Group Theory Considerations}

By an $SO(6)$ rotation the antisymmetric $\mathbf{r}$-matrix can
always be put in the following canonical form:
\begin{equation}
\label{rmatrix}
 \mathbf{r}\rightarrow \mathbf{r}_{can}=\left(\begin{array}{cccccc}
 &-r_1&&&&\\r_1&&&&&\\&&&-r_2&&\\&&r_2&&&\\&&&&&-r_3\\&&&&r_3&
 \end{array}
 \right) \ .
\end{equation}
where, since the $\mathbf{r}$-matrix has to be an orthogonal
matrix in a subspace in which it is non-degenerate, there are only
four choices for $r_\alpha$:
\begin{equation}\label{rs}
 (r_1,r_2,r_3)\in\{(1,0,0), (1,1,0), (1,1,1), (1,1,-1)\}.
\end{equation}
Any different choice of ``gauge'' is equivalent to them. For
example, the second solution can be chosen as well to be
$(1,-1,0)$. However, since $SO(6)$ can not be enlarged to be
$O(6)$ as the global symmetry for the $\CN{4}$ SYM, the last two
solutions are {\em not} equivalent. In the next section we will
see that, if the number of the non-vanishing $r_\alpha$'s is one,
two or three, respectively, the corresponding $R$-ball states
maintain $1/2$, $1/4$ or $1/8$ supersummetry.

\subsection{Non-Commutative Solutions at Large $N$}

In color space, the scalar fields $\phi_s$ are $N$-by-$N$
matrices. The commutative ansatz (\ref{BPScC}) has been used
before in ref.~\cite{Ber04a} to define a holomorphic normal matrix
model. The above approach allows us to consider more sophisticated
$R$-balls by going beyond this ansatz but still having good
control. The simplest case is that in the large $N$ limit, the
commutators $[\phi_s,\phi_t]$ are proportional to the unit matrix
in color space: \beq{NC}
 [\phi_s,\phi_t]= i \frac{\theta_{st}}{R^4},
\eeq where the non-commutative (NC) parameters $\theta_{st}$ are
anti-symmetric and of the dimension of length squared. Other
ansatz or conditions in the last subsections are unchanged, except
perhaps the Gauss's law.

By taking derivative, one has
\begin{equation}
 [D_0\phi_s,\phi_t]+[\phi_s,D_0\phi_t]=
i \frac{\dot{\theta}_{st}}{R^4}.
\end{equation}
By Eq.~(\ref{BPScD0}) and the antisymmetry of $\mathbf{r}$, we get
the equation of motion for the NC parameter:
\begin{equation}
 \dot{\Theta}=\pm [\mathbf{r},\Theta] \ ,
\end{equation}
where $\Theta$ is the matrix $(\theta_{st})$. For simplicity, in
this paper we will only consider the case with $\dot{\Theta}=0$,
or equivalently $[\mathbf{r},\Theta]=0$. Accordingly, in the same
basis for (\ref{rmatrix}), $\Theta$ can be put in a canonical
form:
\begin{equation}\label{thetamatrix}
 \Theta\rightarrow \Theta_{can}=\left(\begin{array}{cccccc}
 &-\theta_1&&&&\\\theta_1&&&&&\\&&&-\theta_2&&\\&&\theta_2&&&\\
 &&&&&-\theta_3\\&&&&\theta_3&
 \end{array}
 \right).
\end{equation}
In addition, one must keep in mind that, for any $r_\alpha=0$, the
corresponding $\theta_\alpha=0$. Moreover, the Gauss's law
requires
\begin{equation}\label{Gauss}
 [D_0\phi_s,\phi_s]=
\pm {i\over R^5}{\rm Tr}_{6\times 6}(\mathbf{r}\Theta)=0;
\end{equation}
in the canonical forms for $\mathbf{r}$ and $\Theta$, we have
\begin{equation}\label{Gauss0}
 \sum\limits_{\alpha=1}^3\theta_\alpha r_\alpha=0.
\end{equation}
There are two straightforward implications: (i) If there is only
one $r_\alpha\neq 0$, then the deformation (\ref{NC}) with
constant $\theta$ violates the Gauss's law. (ii) If two
$r_\alpha\neq 0$ and we choose the gauge $r_1=-r_2=1$, then
$\theta_1=\theta_2$ and $\theta_3=0$. In the second half of the
next section, we will look for $R$-balls in the case (ii).

The deformation (\ref{NC}) is self-consistent only if
$N\to\infty$. Then various sums over colors in the Hamiltonian
become divergent. To define a well-behaved large $N$ limit, one
needs to properly redefine the the trace, $\rm{Tr}$, and the
fields by rescaling them with some negative powers of $N$. In
fact, let us examine the relevant terms in $H$:
\begin{equation}
 H=2\pi^2R^3{\rm Tr}({\phi_s^2\over R^2}
-{g^2\over 4}[\phi_s,\phi_t]^2) \ ,
\end{equation}
where the factor $2\pi^2R^3$ is the volume of $S^3$ with radius
$R$. Now we perform the standard large $N$ trick, by redefining
$\phi_s=\sqrt{N}\tilde{\phi}_s$, $\lambda=g^2N$. Then we extract
the scale from $\Theta$, write it as $\theta$ and identify the
dilatation operator with
\begin{equation}\label{Dil}
 \Delta = {R\over 2\pi^2} H =
N\mathrm{Tr}({\theta\over R^2}\varphi_s^2 -{\lambda\over
4}({\theta\over R^2})^2[\varphi_s,\varphi_t]^2),
\end{equation}
where $\varphi_s=R^2\tilde{\phi}_s/\sqrt{\theta}$ and the
pre-factor $N$ plays the role of $\hbar^{-1}$. The concrete color
space ``renormalization'' scheme will be specified for different
choices of $\Theta$. We will consider two examples below.

First, if $\theta=|\theta_1|\neq 0$ and $\theta_{2,3}=0$, then we
set the following order estimation: ${\rm Tr}\varphi_s^2\sim
\mathcal{O}(N^2)$ while $[\varphi_s,\varphi_t]\sim {\cal O}(1)$,
which is in consistence with (\ref{NC}). Then it is natural to
take $N^2\theta/R^2$ fixed, written as $c_1$; the dilatation
operator becomes
\begin{equation}\label{DE1}
 \Delta=N\Bigl(c_1{\mathrm{Tr}\over N^2}(\varphi_s^2)
 -{\lambda c_1^2\over 4}{\mathrm{Tr}\over N^4}
 ([\varphi_s,\varphi_t]^2)\Bigr).
\end{equation}
Notably, the symbols $\rm{Tr}/N^2$ play the role of the
regularized traces. Moreover, the ratio of the quartic to the
quadratic terms in Eq.~(\ref{DE1}) is of order $\lambda c_1
N^{-3}$. We define the 't Hooft coupling $\lambda$ and $c_1$ to be
the physical parameters independent of the cutoff $N$, then the
quartic terms becomes irrelevant. Meanwhile, the BPS bound is
again saturated.

For the second example, let us consider the case with
$\theta_1=\theta_2$ with absolute value $\theta$ and $\theta_3=0$.
The solution to Eq.~(\ref{NC}) has a direct product structure (see
eqs. (\ref{19b}) and (\ref{SS}) in next section), and two
regulators $N_1$ and $N_2$ are needed such that $N_1N_2=N$. Then
the order estimation is set to be ${\rm Tr}\varphi_s^2\sim
\mathcal{O}(N(N_1+N_2))$. In this case, the physical parameter is
given by $c_2=N(N_1+N_2)\theta/R^2$ and the physical dilatation is
\begin{equation}\label{DE2}
 \Delta=N\Bigl(c_2{\mathrm{Tr}(\varphi_s^2)\over {N(N_1+N_2)}}
 -{\lambda c_2^2\over 4}{\mathrm{Tr}
([\varphi_s,\varphi_t]^2)\over N^2(N_1+N_2)^2}\Bigr).
\end{equation}
This time, $\rm{Tr}/N(N_1+N_2)$ serve as the regularized trace;
the ratio of the quartic to the quadratic terms in Eq.~(\ref{DE2})
is of order $\lambda c_2/N(N_1+N_2)^2$ with $c_2$ and $\lambda$ to
be fixed, independent of $N$. Note that in both (\ref{DE1}) and
(\ref{DE2}), one should {\em not} further absorb $c_1$ and $c_2$
into another redefinition of fields, because this would change the
universality classes of the model.

In summary, with the non-commutative ansatz (\ref{NC}), the BPS
bound~(\ref{BE}) is saturated up to an error that vanishes in a
large $N$ limit. We will refer this type of configurations as {\em
almost-BPS}. Allowing us to see this new possibility is a
significant advantage of our approach.

\section{$R$-Ball Solutions with Unbroken SUSY}

With only bosonic backgrounds, unbroken supersymmetry requires
$\delta_\zeta\psi=0$. In this section we will find the BPS
$R$-ball configurations that also preserve part of supersymmetry,
i.e. satisfy the conditions (from (\ref{susy}))
\beq{9}0=\delta_{L}\psi_a&=&
-i\gamma^0(\alpha^i\dot{X}_{ia}+i\gamma_5\beta^j\dot{Y}_{ja})
\zeta_{L} +\frac{1}{R}(\alpha^iX_{ia}-i\gamma_5\beta^jY_{ja})
\gamma_5\gamma^0\zeta_{L} \nn \\
&&+\frac{ig}{2}\ep^{ijk}(\alpha^k[X_i,X_j]_a+\beta^k[Y_i,Y_j]_a)
\zeta_{L}
+g[X_i,Y_j]_a\alpha^i\beta^j\gamma_5\zeta_{L}, \nn \\
0=\delta_{R}\psi_a&=&
-i\gamma^0(\alpha^i\dot{X}_{ia}+i\gamma_5\beta^j\dot{Y}_{ja})
\zeta_{R} -\frac{1}{R}(\alpha^iX_{ia}-i\gamma_5\beta^jY_{ja})
\gamma_5\gamma^0\zeta_{R} \nn \\
&&+\frac{ig}{2}\ep^{ijk}(\alpha^k[X_i,X_j]_a+\beta^k[Y_i,Y_j]_a)
\zeta_{R} +g[X_i,Y_j]_a\alpha^i\beta^j\gamma_5\zeta_{R}, \eeq with
$\zeta_{L,R}$ not all vanishing. Here $a=0,1,2,...,N^2-1$ are the
indices of the adjoint $U(N)$ representation. A representation of
the algebra (\ref{3}) is chosen to be \beq{14}
&&\alpha_1=i\sigma_2\times\sigma_1,\hspace{0.6in}
\alpha_2=-i\sigma_2\times\sigma_3,\hspace{0.6in}
\alpha_3=i{\bf 1}_{2\times 2}\times\sigma_2,\nn \\
&&\beta_1=-i\sigma_1\times\sigma_2,\hspace{0.6in}
\beta_2=-i\sigma_2\times{\bf 1}_{2\times 2},\hspace{0.6in}
\beta_3=i\sigma_3\times\sigma_2. \eeq

In this section, we will not only find classical solutions
satisfying these conditions, but also count the moduli of the
solutions of a given type. A clear understanding of the moduli of
the solution space is crucial for the collective coordinate
quantization we are going to apply in the next section. This is
because the moduli form the configuration space of the collective
coordinates for a given type of solutions, and the collective
coordinate quantization heavily exploits the knowledge of the
moduli space. To avoid overcounting, one needs to be careful:
$SO(6)$ inequivalent configurations may be gauge equivalent, since
the scalars carry both global $SO(6)$ and local $U(N)$ degrees of
freedom, which may be entangled in the moduli counting.

\subsection{Commutative $R$-Balls and Their Moduli}

We first consider the commutative ansatz (\ref{BPScC}) with only
$r_1\neq 0$ and $r_2=r_3=0$ in the canonical form (\ref{rmatrix}).
In this case, we need to consider only one pair of scalar fields
$X=X_1$ and $Y=Y_1$. Then the supersymmetry condition~(\ref{9})
reduces to two systems of linear equations:
\beq{15}\left\{\begin{array}{l}
G_a\zeta_1+(K_a-F_a)\zeta_4=0, \\
G_a\zeta_2-(K_a+F_a)\zeta_3=0, \\
(K_a+F_a)\zeta_2+G_a\zeta_3=0, \\
(K_a-F_a)\zeta_1-G_a\zeta_4=0,
\end{array}\right.\quad\quad \mbox{and} \quad\quad
\left\{\begin{array}{l}
G_a\td{\zeta}_1+(\td{K}_a-\td{F}_a)\td{\zeta}_4=0, \\
G_a\td{\zeta}_2-(\td{K}_a+\td{F}_a)\td{\zeta}_3=0, \\
(\td{K}_a+\td{F}_a)\td{\zeta}_2+G_a\td{\zeta}_3=0, \\
(\td{K}_a-\td{F}_a)\td{\zeta}_1-G_a\td{\zeta}_4=0,
\end{array}\right.
\eeq
where $\zeta_A,\;\td{\zeta}_A$ $(A=1,2,3,4)$ are the ``left'' and
``right'' Majorana conformal Killing spinors and
\beq{16}G_a=g\gamma_5[X,Y]_a,\hspace{0.3in}
K_a&=&-i\gamma^0\dot{X}_a+\gamma_5\gamma^0X_a/R,\hspace{0.5in}
F_a=\gamma^0\gamma_5\dot{Y}_a-i\gamma^0Y_a/R, \nn \\
\td{K}_a&=&-i\gamma^0\dot{X}_a-\gamma_5\gamma^0X_a/R,\hspace{0.5in}
\td{F}_a=\gamma^0\gamma_5\dot{Y}_a+i\gamma^0Y_a/R. \eeq The
vanishing determinant of each system leads to the conditions for
unbroken supersymmetry: \begin{equation}\label{17}
(|-i\dot{Z}_a+R^{-1}Z_a|^2-(\frac{g}{2}[Z,Z^\dag]_a)^2)
(|i\dot{Z}_a+R^{-1}Z_a|^2-(\frac{g}{2}[Z,Z^\dag]_a)^2)=0,
\end{equation}
with $Z=X+iY$. It is easy to see that when $[X,Y]=0$, i.e.
$[Z,Z^\dag]=0$, the above equations are reduced to the BPS
condition (\ref{BPScD0}) in the $A_0=0$ gauge. The latter is
easily solved, resulting in \beq{18} Z=e^{\pm it/R}A \ , \eeq with
$A$ any $N\times N$ time-independent normal matrices:
$[A,A^\dag]=0$. Inserting the solutions (\ref{18}) back into
Eq.~(\ref{15}), it is easy to verify that there are $16$
supercharges. Hence the solutions (\ref{18}) are $1/2$-BPS
backgrounds.

Other examples can be worked too. It can be verified that if the
number of non-zero $r_\alpha$'s in the canonical form
(\ref{rmatrix}) is $\gamma$, then the fraction of unbroken
supersymmetry is $1/2^\gamma$. An important question is how many
commutative BPS $R$-balls there are. This is the problem of
counting the moduli of such solutions, which we will address here.

With the canonical form (\ref{rmatrix}) of $\mathbf{r}$, one may
define complex scalars as
$Z_\alpha=\phi_{2\alpha-1}+i\phi_{2\alpha}$ $(\alpha=1,2,3)$. Then
the equations (\ref{BPScDi}), (\ref{BPScD0}) and (\ref{BPScC})
that completely determine the commutative BPS $R$-balls can be
recast into the form:
\begin{eqnarray}
\label{NED0}
 \dot{Z}_\alpha=i{r_\alpha\over R}Z_\alpha, \qquad
 [Z_\alpha,Z_\beta]=[Z_\alpha,Z_\beta^\dag]=0.
\end{eqnarray}
The first equation can be solved by $Z_\alpha=A_\alpha
e^{ir_\alpha t/R}$ with $A_\alpha$ time-independent $N$-by-$N$
matrices. The second equation indicates that $A_\alpha$ can be put
in the form
\begin{equation}\label{CCg}
 A_\alpha=U^\dag a_\alpha U
\end{equation}
with $a_\alpha$ (the eigenvalue matrix) diagonal: $a_\alpha =
diag(a_{\alpha1,1} a_{\alpha,2},\cdots, a_{\alpha,N})$, and $U$
unitary. Since all diagonal $U$'s give rise to the same $A_\alpha$
when $a_\alpha$ is diagonal, the solution space for a particular
$R$-charge generator $\mathbf{r}$ is given by
\begin{equation}
 \mathcal{M}_{\mathbf{r},N}=\{(a_\alpha,U)\}
=\mathbf{C}^{\gamma N}\times U(N)/U(1)^N \ .
\end{equation}
(recall that $\gamma$ is the number of non-zero $r_\alpha$.)
Furthermore, recall the BPS bound (\ref{BE})
\begin{equation}
 Q_\mathbf{r}=\pm {1\over 2\pi^2 R}\int\limits_{S^3}
\sum\limits_\alpha \mathrm{Tr}(Z^\dag_\alpha
 Z_\alpha)=\pm R^2\sum\limits_{\alpha,i}|a_{\alpha,i}|^2\ ,
\end{equation}
where we have absorbed a volume factor $2\pi^2$ into the
$R$-charge. Thus, an $R$-charge sector is a class of $R$-ball
solutions with the same $R$-charge $Q_\mathbf{r}$ associated with
generator $\mathbf{r}$. Once the $R$-charge generator is
specified, the moduli space $\mathcal{M}_{\mathbf{r},N}$ is
divided into different $R$-charge sectors, and in each sector the
value of the dilatation $\Delta$ defined by Eq.~(\ref{Dil}) is
fixed by the BPS condition to be $\Delta=Q_\mathbf{r}$.

As an example, for the $1/2$ BPS $R$-charge sector with $r_1=1$,
$r_2=r_3=0$, the $R$-charge $Q$ is given $Q=R^2\sum\limits_i
|a_i|^2$. Similarly, each commutative $1/2^\gamma$ BPS $R$-charge
sector with a canonical $\mathbf{r}$ defines a sphere $S^{2\gamma
N-1}$ in eigenvalue space $\mathbf{R}^{2\gamma N}$.

\subsection{Non-commutative $R$-Balls and Their Moduli}

Now we consider the non-commutative ansatz (\ref{NC}), which will
leads to new solutions in large $N$. We have shown that a single
non-vanishing complex scalar $Z$, non-commutative, almost-BPS
background violates the Gauss's law (\ref{Gauss}), since
$[\dot{Z}, Z^\dag]+[\dot{Z}^\dag, Z] \neq 0$. So we need, at
least, to turn on two complex scalars:
$Z_1=X_1+iY_1,\;Z_2=X_2+iY_2$. Based on the discussions in the
previous section, we should pick $r_1=1$ and $r_2=-1$ for the
$R$-charge generator. With this choice we have the solutions
\beq{19b} Z_1={1\over R^2}e^{it/R} A_1,~~Z_2={1\over R^2}e^{-it/R}
A_2, \eeq where the time-independent matrices $A_1$ and $A_2$ obey
a two-dimensional Heisenberg algebra: \beq{19c}
&&[A_1,A_1^\dag]=2\theta_1,\hspace{1in}
[A_2,A_2^\dag]=2\theta_2, \nn \\
&&[A_1,A_2]=[A_1,A_2^\dag]=[A_2,A_1^\dag]=[A_1^\dag,A_2^\dag]=0,
\eeq with $\theta_1=\theta_2$, $|\theta_1|=\theta$. The matrices
$A_1$ and $A_2$ span two orthogonal non-commutative planes and the
Gauss's law constraint is satisfied because $\theta_1=\theta_2$.

Now let us count the number of unbroken supersymmetries via
solving the supersymmetry variation condition~(\ref{9}), that in
this case reduces to:
\beq{M1}\left\{\begin{array}{l}
(G_{11}-G_{22})_a\zeta_1-(K_2+F_2)_a\zeta_3+(K_1-F_1)_a\zeta_4=0, \\
-(G_{11}-G_{22})_a\zeta_2+(K_1+F_1)_a\zeta_3+(K_2-F_2)_a\zeta_4=0, \\
(K_2+F_2)_a\zeta_1-(K_1+F_1)_a\zeta_2-(G_{11}+G_{22})_a\zeta_3=0, \\
-(K_1-F_1)_a\zeta_1-(K_2-F_2)_a\zeta_2+(G_{11}+G_{22})_a\zeta_4=0,
\end{array}\right.
\eeq in which \beq{M2}G_{ij}=g\gamma_5[X_i,Y_j],\hspace{0.3in}
K_i&=&-i\gamma^0\dot{X}_i+\gamma_5\gamma^0X_i/R,\hspace{0.5in}
F_i=\gamma^0\gamma_5\dot{Y}_i-i\gamma^0Y_i/R \eeq for $i,j=1,2$,
with color indices suppressed. Since the solution~(\ref{19b}) and
the commutation relation~(\ref{19c}) lead to
\beq{M3}G_{11}-G_{22}=K_1+F_1=K_2-F_2=0, \eeq the solution
of~(\ref{M1}) is $\zeta_{1,3,4}=0$, leaving $\zeta_2$ the only
surviving Killing spinor. Similar result holds for
$\tilde{\zeta}_A$. So among all 32 components of Killing spinors,
only a quarter of them can be nonzero and linearly independent.
Therefore, the classical non-commutative configurations
(\ref{19b}) with (\ref{19c}) preserve eight supersymmetries. This
class of 1/4 BPS backgrounds has not been discovered before in the
literature.

The moduli space, $\mathcal{M}_{\mathbf{r},\Theta}$, for this
class of non-commutative $R$-balls is qualitatively different from
that of the commutative BPS $R$-balls. To start, we rewrite the
non-commutative ansatz (\ref{NC}) in the exponential form:
\begin{equation}\label{NCE}
 \exp(i\phi_su^s)\exp(i\phi_tv^t)
=e^{-\theta_{st}u^sv^t/R^4} \exp(i\phi_tv^t)\exp(i\phi_su^s) \ ,
\end{equation}
where $u^s$ and $v^t$ are two vectors in $R^6$. By the celebrated
Stone-von Neumann theorem, any solutions to (\ref{NCE}) are
unitarily equivalent. Then we focus on the case of (\ref{19b}).
Because $r_1$ and $r_2$ is now gauged to be $1$ and $-1$, the sign
of $\theta_1=\theta_2$ matters; so there are actually two
different solutions $(\theta,\theta)$ and $(-\theta,-\theta)$,
where $\theta$ by definition is non-negative. Without losing of
generality, we take $(\theta,\theta)$. In this case,
\begin{equation}\label{SS}
 A_\alpha = \sqrt{2\theta} U^\dag a_\alpha U,
~a_1=a\times {\bf 1}, ~a_2={\bf 1}\times a\ ,
\end{equation}
where $a$ is a standard matrix representation in quantum mechanics
\beq{32}
a=\left(\begin{array}{ccccc} 0&\quad 1&&& \\
&0&\quad\sqrt{2} && \\ &&\ddots&\quad\sqrt{3} & \\ &&&\ddots &
\ddots
\end{array} \right) \,
\eeq and $U$ is an infinite-dimensional unitary matrix. So
$\mathcal{M}_{\mathbf{r},\Theta}$ is the product of
$\mathbf{R}^+=\{\sqrt{\theta}\}$ and an infinite special unitary
group ``$SU(\infty)$'', loosely speaking. However, the difference
with the commutative half-BPS case is that only a $U(2)$ subgroup
in this $SU(\infty)$ will be considered as dynamical variables,
with the rest being gauge degrees of freedom. This $U(2)$ is
generated by the Schwinger representation,
\begin{equation}\label{SchR}
 \mathbf{L}_\mu=a_\alpha^\dag (\sigma_\mu)^\alpha_{\beta}a^\beta,
\end{equation}
with $\mu=0,1,2,3$, $\sigma_0=1$. It is easy to show
$\mathbf{L}_{1,2}$ do not contribute to energy in (\ref{DE2}).
According to the above-mentioned analysis, we reparameterize this
type of $R$-balls as
\begin{equation}
 A_\alpha = \sqrt{2\theta}e^{i\varphi_\alpha} V^\dag a_\alpha V,
 \qquad \alpha=1,2,
\end{equation}
with $V\in SU(\infty)/U(2)$. $\mathbf{L}_0$ generates the
translation in the direction of $\varphi_1+\varphi_2$ while
$\mathbf{L}_3$ generates the translation in the direction of
$\varphi_1-\varphi_2$. And in the later treatment,
$\sqrt{\theta}$, $\varphi_\alpha$ are dynamical while $V$ plays
the role of gauge degrees of freedom.

Now let us regularize $a_1$ and $a_2$ by truncating (\ref{32}) to
estimate the $N$-dependence of the $R$-charge. The first factor in
the direct product (\ref{SS}) is regularized by the upper-left
$N_1$-by-$N_1$ block, while the second factor by the upper-left
$N_2$-by-$N_2$ block. Accordingly, we have $N=N_1N_2$. The
$R$-charge calculated from Eq.~(\ref{DE2}) (with the quartic term
safely thrown away) is given by
\begin{equation}\label{ES}
Q = Nc_2 = N^2(N_1+N_2){\theta\over R^2}.
\end{equation}
The prefactor $N$ is familiar in any quadratic quantities in large
$N$ field theories. Now the $R$-ball sector is specified by the
``renormalized $R$-charge'' $c_2$, instead of the bare ratio
$\theta/R^2$.

\section{Quantization of the Commutative Half BPS Sector}

The classical $R$-balls have continuous values for the $U(1)$
$R$-charge. It is necessary to quantize the $R$-balls in order to
get a discrete spectrum for the $R$-charge. In this section, we
will quantize the $R$-balls by both collective coordinate
quantization \cite{CL75,FLS76,ZZDW80} and canonical quantization.
In the commutative half BPS sector we will show that both
quantization reproduce the previous results obtained by the matrix
model approach \cite{Ber04a}, but our treatment will shed new
light on several important aspects of physics. In particular, we
can explicitly exhibit the origin of the Landau levels in the
present text, so as to make the connection of the BPS dynamics
with the quantum Hall effect meaningful and substantial.

\subsection{Collective Coordinate Quantization}

We have seen the time-independent matrix $A$ in our
solution~(\ref{18}) can be put in the form
\begin{equation}\label{CC}
 A= U^\dag a U,
\end{equation}
where $a=diag(a_1,a_2,\ldots, a_N)$ and $U\in U(N)/U(1)^N$.
Consequently, the collective coordinate space for solution
(\ref{18}) is identified to be
\beq{moduli1}
\mathcal{M}_{\mathbf{r},N}=\left(\mathbf{C}^N/S_N\right)\times
\left(U(N)/U(1)^N\right).
\eeq
Here $S_N$ is the symmetric group of degree $N$. As we will see,
the collective coordinate quantization on this moduli space will
lead to a Hilbert space including non-BPS quantum states, while
the quantum BPS states form only a subspace.

The collective coordinate quantization was originally developed
for (topological and non-topological) solitons in scalar field
theory. Classically the internal conserved observables of the
solitons are generated by rotating the collective coordinates of a
static solution in internal space. So to quantize the value of the
internal observables, naturally one needs to turn the collective
coordinates into quantum dynamical variables. In the present case,
we promote the variables $a$ and $U$ in Eq.~(\ref{CC}) to
dynamical variables:
\begin{equation}\label{Q}
 a\rightarrow a(t), ~~ U\rightarrow U(t).
\end{equation}
(Alternatively, we may absorb the exponential factor $e^{-it/R}$
into the diagonal part with $a(t) \to a(t) e^{it/R}$. We will not
use this convention however.) Recall that Eq.~(\ref{CC}) is in the
$A_0=0$ gauge. There is a residue global $U(N)$ symmetry in this
gauge, and because of the original color gauge symmetry in the SYM
we need to impose the Gauss's law constraints.

Substituting Eq.~(\ref{CC}) with (\ref{Q}) into the original
Lagrangian of SYM, we get the Lagrangian for the  collective
coordinates $a(t)$, $U(t)$:
\begin{equation}\label{CCL}
 L={\rm Tr}\{\dot{a}\dot{a}^\dag
 -{i\over R} (a^\dag\dot{a}-a\dot{a}^\dag)
 -{1\over 2}[a, \omega][a^\dag, \omega]\} \,
\end{equation}
where $\omega:=i\sqrt{2}\dot{U}U^\dag$, $\omega^\dag=\omega$.
Because $U\in U(N)/U(1)^N$, $\omega^i_i=0$ for $i=1,2,\ldots, N$.
In terms of the matrix elements, the Lagrangian (\ref{CCL}) reads
\begin{equation}
\label{CCLm}
 L=\sum\limits_{i=1}^N(|\dot{a}_i|^2-{i\over R}
 (a_i^\ast\dot{a}_i-a_i\dot{a}_i^\ast)
 + {1\over 2}\sum\limits_{i\neq j}|a_i-a_j|^2
 \omega^i_j\omega^j_i.
\end{equation}

The first two terms in Eq.~(\ref{CCLm}) define a standard Landau
problem for $N$ particles, with cyclotron frequency $1/R$. The
origin of the ``magnetic field'' is due to the rotation
$e^{-it/R}$ in the $R$-ball solution that generates $R$-charge.
The last term is in the standard form for a top rotating in a
homogeneous space with symmetry $G$, with $I_{AB}$ the {\em
inertia tensor} and $\omega^A$ the {\em angular velocities} taking
values in the Lie algebra of $G$. In the present case, the group
$G$ is $U(N)$. The inertia tensor is diagonal with element
$I^{ii}_{jj}$ given by $|a_i-a_j|^2$. The canonical momenta are
\begin{equation}\label{AM}
 J^i_j={\partial L\over\partial\omega^j_i}
 =\left\{\begin{array}{ll} |a_i-a_j|^2\omega^i_j,
   &i\neq j;\\
 0,&i=j\end{array}\right\} ,
\end{equation}
with the Poisson structure
\begin{equation}\label{P}
 \{J^i_j,J^k_l\}_{P.B.}=\delta^i_lJ^k_j-\delta^k_jJ^i_l \ .
\end{equation}

Observe that
\begin{equation}\label{Fermi}
 a_i\neq a_j,~~~\forall i\neq j~\Longleftrightarrow
 det({\partial^2L\over\partial\omega^i_j\partial\omega^k_l})
=|\Delta(a)|^2\neq 0 \ ,
\end{equation}
where $\Delta(a)$ is the van DeMonde determinant for
$(a_1,a_2,\ldots,a_N)$:
\begin{equation}
 \Delta(a)=\prod\limits_{i<j}(a_i-a_j).
\end{equation}
So the Hessian for (\ref{AM}) is nonsingular and, therefore, the
Hamiltonian is well-defined {\em only} in the subspace of moduli
without coinciding eigenvalues. We will do canonical quantization
on this subspace with the Hamiltonian
\begin{equation}\label{H}
 H=H_0 +\sum\limits_{i\neq j}{J^i_jJ^j_i\over 2|a_i-a_j|^2},
\end{equation}
where
\begin{equation}\label{H0}
 H_0=\sum\limits_{i=1}^N(p_i +{ia_i^\ast\over R})
     (p_i^\ast-{ia_i\over R}) \ ,
\end{equation}
and the canonical momenta are given by
$p_i=\dot{a}_i^\ast-ia_i^\ast/R$, $p_i^\ast = \dot{a}_i+ia_i/R$.
The Hamiltonian in (\ref{H}) is a {\em generalized}
Calogero-Sutherland model for $U(N)$-spin \cite{Poly97} coupled to
a constant magnetic field. By ``generalized'', it meant that the
variables $a_i$ are complex instead of real numbers. This
difference will dramatically change the physics at the quantum
level.

To quantize the system, we first promote the Poisson brackets
(\ref{P}) to the commutation relations for $su(N)$ Lie algebra:
\begin{equation}\label{P0}
 [J^i_j,J^k_l]=i(\delta^i_lJ^k_j-\delta^k_jJ^i_l) \ .
\end{equation}
Classically the Gauss's law $[Z^\dag,D_0Z]+[Z,D_0Z^\dag]=0$ on the
moduli space reads
\begin{equation}\label{GL}
 [a^\dag,[a,\omega]]+[a,[a^\dag,\omega]]=0,
\end{equation}
or equivalently, in terms of the $U(N)$ angular momenta,
\begin{equation}\label{GL0}
 J^i_j=0.
\end{equation}
At the quantum level, the Gauss's law (\ref{GL0}) is promoted to
the constraints on the physical states:
\begin{equation}\label{GL1}
 J^i_j|phys\rangle = 0.
\end{equation}

To see the meaning of the constraints, we introduce a coordinate
representation: $|phys\rangle\rightarrow \psi(a,U)$. Then $J^i_j$
are represented by the {\em right-invariant vector fields} on the
$U(N)$ group manifold that generate left translations:
\begin{equation}
 J^i_j=-iU^i_k{\partial\over\partial U^j_k};~~
 \left(1+\epsilon^i_jJ^j_i\right)\, f(U)= f\left((1-i\epsilon)U\right).
\end{equation}
The Lie algebra relation (\ref{P0}) is readily to verify.
Moreover, it is obvious that $J^i_i\psi=0$. The Gauss's law
(\ref{GL1}) is equivalent to $\psi(a,U)=\psi(a)$. Namely, the
wavefunction of physical states are independent of the coordinates
$U$. So we will consider only physical states in the form
$\psi(a)$. Thus the physical degrees of freedom are reduced to the
diagonal elements $a_i$, giving rise to the many-body
interpretation of the quantum states. As we have seen from
(\ref{Fermi}), a Hamiltonian formalism is well defined only on a
subspace of the moduli with $a_i$ all unequal. The reduced moduli
space for $a_i$'s is then $\{\mathbf{C}^N-D\}/S_N$, where $D$ is
the set of points in $\mathbf{C}^N$ with coinciding coordinates.
The fundamental group of this reduced moduli space is known to be
the braid group of $N$-particles that classifies the quantum
statistics in two dimensions \cite{Wu84}. This is the origin of
the emergence of non-trivial statistics, including fermions, after
quantizing scalar (bosonic) field configurations.

Finally, we consider the Hamiltonian in the subspace of physical
states. There is a nontrivial measure in defining the inner
product in the physical Hilbert space. This measure can be viewed
as the Faddeev-Popov measure due to gauge fixing. In fact, in the
space of all normal matrices, the measure is
$dAdA^\dag=d\mu(U)dada^\dag|\Delta(a)|^2$, where $d\mu(U)$ is the
descended Haar measure at point $U$ on the coset space
$U(N)/U(1)^N$. By integrating out the unphysical ``angular part''
$d\mu(U)$, the inner product of two physical states is given by
\begin{equation}
 \langle \phi|\psi\rangle = \int \prod_i
 da_ida_i^\dag|\Delta(a)|^2\phi(a)^\ast\psi(a).
\end{equation}
Then the Hamiltonian acting on the wavefunction $\psi$ is
identified with $H_0$ in (\ref{H0}) with the measure factor taking
into account:
\begin{equation}\label{H0Q}
 H=\sum\limits_{i=1}^N{1\over |\Delta|^2}(p_i +{ia_i^\ast\over
 R})|\Delta|^2(p_i^\dag-{ia_i\over R}) \ ,
\end{equation}
in which
\begin{equation}
 p_i=-i{\partial\over\partial a_i}, ~
 p_i^\dag = -i{\partial\over\partial a_i^\ast}.
\end{equation}
A statistical interaction appears in $H$ because of the nontrivial
measure factor $|\Delta(a)|^2$. Similar to the one-dimensional
case, this statistical interaction can be absorbed by a
redefinition of wavefunction:
\beq{flux} \psi(a)\rightarrow \Psi(a)=:
\Delta(a)\psi(a)\ .
\eeq
The Hamiltonian acting on the wavefunction $\Psi$ is given by
\begin{equation} \label{LL}
 \mathbf{H}=\sum\limits_{i=1}^N {1\over \Delta^\ast}
(p_i+{ia_i^\ast\over R}) \Delta^\ast\Delta (p_i^\dag-{ia_i\over
R}){1\over\Delta} =\sum\limits_{i=1}^N (p_i+{ia_i^\ast\over R})
(p_i^\dag-{ia_i\over R}),
\end{equation}
which has the same form of $H_0$ due to the facts that
$[p_i^\dag,\Delta]=0$, $[p_i,\Delta^\ast]=0$. In deriving the
second equality, we have explored the holomorphy of the factor
$\Delta$. The transformation (\ref{flux}) has the effect of
attaching a statistical flux \cite{Wu84-2} to each particle in two
dimensions, to turn the original bosons into fermions \cite{Wu84}.
So the Hamiltonian (\ref{LL}) describes a free fermion system in a
magnetic field. (Note that the above treatment is a bit more
sophisticated than that of Berenstein's hermitian matrix toy model
\cite{Ber04a}, because our $Z$ is not hermitian.)

The ground states of the Hamiltonian~(\ref{H0Q}) are determined by
the following first-order equations
\begin{equation}\label{LLL}
 ({\partial\over\partial a_i^\ast}+{a_i\over R})\psi=0, ~~
 i=1,2,\ldots,N.
\end{equation}
For any $i$, the solution to Eq.~(\ref{LLL}) is a lowest Landau
level (LLL): $\phi_m(a_i)=a_i^m e^{-|a_i|^2}/\sqrt{m!}$ for
$m=0,1,\ldots$; and the many-body ground states are the
symmetrization of the LLL's for all $i$ \beq{wf}
\psi_{G}=\mathcal{C}_N \sum_{sym} \prod_i \phi_{m_i}(a_i), \eeq
where $\mathcal{C}_N$ is the normalization factor. Accordingly,
the ground states for (\ref{LL}) are given by
\beq{wff}\Psi_{G}=\Delta(a)\psi_{G}.
\eeq
Thus the quantum half BPS states are identified with the ground
states, which are infinitely degenerate. Excited states come from
higher Landau levels, and generically are non-BPS. (Though the
many-body system is a free system, there are statistical
correlations coming from the measure.) The gap between the LLL and
excited states is of order $1/R$. So in the small-$R$ limit, the
quantum states will be projected down to the LLL, i.e. the half
BPS sector.

\subsection{Canonical quantization}

In this subsection, we are going to show that a direct application
of canonical quantization to the matrix elements of $Z$ for the
commutative half BPS $R$-balls can reproduce the results of
collective coordinate quantization. Recall the classical system to
be quantized:
\begin{equation}\label{1/2}
 \dot{Z}=iZ,~[Z,Z^\dag]=0,
\end{equation}
where we have set $R=1$ so that all quantities are dimensionless.
The canonical momenta are identified to be $P=\dot{Z}^\dag$ for
this first-order system. We promote each matrix element of $Z$ to
an operator, and impose the canonical commutation relations
$[Z^i_j,P^k_l]=i\delta^i_l\delta^k_j$. Using the first equation in
(\ref{1/2}), we have
\begin{equation}
 [Z^i_j,Z^{\dag k}_l]=-\delta^i_l\delta^k_j.
\end{equation}
The second equation in (\ref{1/2}) now can not hold for operators;
so we impose it as constraints on physical states. More
explicitly, we introduce
\begin{equation}
 L^i_j:=Z^i_kZ^{\dag k}_j-Z^k_jZ^{\dag i}_k.
\end{equation}
Note that they are automatically traceless: $\sum_i L^i_i=0$. It
is easy to check that they generate the $su(N)$ Lie algebra:
\begin{equation}
 [L^i_j,L^k_l]=\delta^k_jL^i_l-\delta^i_lL^k_j.
\end{equation}
In other words, $L^i_j$ provide the Schwinger oscillator
representation of $su(N)$. In this formalism, the second equation
in (\ref{1/2}) is promoted to the constraints
\begin{equation}
 L^i_j|phys\rangle = 0,
\end{equation}
so the quantum states are $SU(N)$ singlets. For these states,
Gauss's law $([Z,\dot{Z}^\dag]+[Z^\dag,\dot{Z}])|phys\rangle =0$,
and the BPS conditions $(\Delta-Q)|BPS\rangle =0$ are
automatically satisfied because of eqs. (\ref{1/2}), with
\begin{equation}\label{B}
 \Delta=Q={\rm Tr}ZZ^\dag.
\end{equation}
In this scheme, all the physical states saturate the BPS bound
$\Delta$=$Q$ quantum-mechanically, since we started with a
first-order system, equivalent to a LLL system.

The wavefunction of quantum states can be defined in the
coherent-state (complex coordinate) representation by
$|phys\rangle\rightarrow \Psi(Z)$ and $Z^\dag\rightarrow
\partial/\partial Z^T$. (Here the superscript $T$ stands for
matrix transpose.) In fact, this is a Bargmann-Fock representation
in the space of holomorphic functions $\Psi(Z)$ with measure
$d\mu(Z,Z^\dag)=e^{-{\rm Tr}ZZ^\dag}$. Recall that
$Z\partial/\partial Z^T$ generates the left $U(N)$ action on $Z$,
while $Z^T\partial/\partial Z$ the right action:
\begin{equation}
 {\rm Tr}(\epsilon Z{\partial\over\partial
 Z^T})\Psi(Z)=\Psi((1+\epsilon)Z)-\Psi(Z),~~
 {\rm Tr}(\epsilon Z^T{\partial\over\partial
 Z})\Psi(Z)=\Psi(Z(1+\epsilon))-\Psi(Z).
\end{equation}
So $L^i_j$ generate {\em similar} transformations in
$SL(N,\mathbf{C})$ by $\Psi(Z)\rightarrow \Psi(gZg^{-1})$ in the
complex domain and the Gauss's law dictates that
\begin{equation}\label{KEY}
 \Psi(gZg^{-1})=\Psi(Z),~\forall g\in SL(N,\mathbf{C}).
\end{equation}
Note that this $SL(N,\mathbf{C})$ is not a symmetry on the Hilbert
space, but a symmetry on the wavefunctions for physical states.
Eqs.~(\ref{KEY}) and (\ref{B}) are the major results in the
literature on $1/2$ BPS states.

\subsection{Quantum Hall Analogy and Holography}

In the above we have obtained the wavefunction (\ref{wff}) for
half BPS quantum states. It describes a many-body system of $N$
particles, which we call $G$-particles. Here the name $G$ hints
about two features of these particles: their origin in the gauge
(color) degrees of freedom (their number $N$ being the rank of the
gauge group $U(N)$) in SYM and their close relation to {\em
geometry} or {\em gravity} in the holographic dual (see below).

The simplest ground state with the wavefunction (\ref{wff})
corresponds to $m_i=0$ for all $i$. It has the minimal angular
momentum or $R$-charge $N^2/2$. In this case, the wavefunction is
nothing but the Laughlin wavefunction for a quantum Hall droplet
with filling factor $\nu=1$, an incompressible quantum fluid
forming a circular disk. (For an introduction of the QHE for
particle physicists see, e.g., \cite{Wu89}; for that for string
theorists see, e.g., \cite{GMSS05,Susskind01a}.) The general
states described by a wavefunction (\ref{wf}) represent planar
fluid composed of discontinuous components of the form of
concentric rings. If the single particle states have larger enough
angular momenta, they form a free 2D fermion gas in the LLL.

Compared with Berenstein's treatment, our $R$-ball approach has
the advantage that one can see clearly the origin of the emergence
of ``Landau levels'', on which the quantum Hall analogy is based.
Essentially this is due to the rotation (the time-dependent factor
$\exp (-it/R)$ in Eq.~(\ref{CC})) of the $R$-balls that generates
$R$-charge. By now it is well-known that the rotation of a BEC
will lead to the emergence of an effective magnetic field in the
co-moving frame and of the single-particle Landau levels
\cite{Wilkin98}. Numerically it has been shown that with small
filling fractions, even fermionic QH-like states \cite{Wilkin00},
including FQH-like states \cite{Wilkin01}, should appear in
rotating BEC's. Actually the LLL states (but not QHE yet) in a
rotating BEC has been seen experimentally \cite{exp}. What we have
seen above in the half BPS sector in SYM is essentially the same
physics: An $R$-ball in $\CN{4}$ SYM is nothing but a rotating
BEC; and the small radius limit ($R \to 0$ corresponds to rapidly
rotating BEC, which is indeed the lowest Landau level regime in
the real atomic BEC experiments.

To understand the solutions (\ref{18}) from the point of view of
dual IIB superstring theory, we claim that the $G$-particles
satisfying Fermi statistics correspond to the LLM fermions that
are the ``sources'' of the half BPS geometry in a large $N$
limit$^1$ \footnotetext[1]{If $N$ is not large enough, in general
we do not have classical geometry in dual string theory.} in the
LLM's construction\cite{LLM}. The state with minimal $R$-charge
(for fixed $N$) is a circular droplet with a uniform distribution
of the LLM fermions, which is known to be the ``source'' of the
$AdS_5\times S^5$ geometry. The states whose $R$-charge are not
far above $N^2/2$ can be viewed as a few $G$-particles excited a
bit outside the droplet, corresponding to a few giant graviton
excitations in $AdS_5$. The general discontinuous fluid states of
$G$-particles that form concentric rings correspond to general
LLM's half BPS geometries seeded by concentric ring-like
distribution of LLM fermions. Finally, generically not every
possible state of the $G$-particle gas correspond to a classical
geometry.

\section{Quantization of Non-commutative $1/4$ BPS Sector}

The success in the last section in making the connection of half
BPS $R$-balls to the quantum Hall effect meaningful substantially
encourages us to proceed to examine the non-commutative quarter
BPS R-balls and to confirm the appearance of FQH-like states after
quantization.

\subsection{Collective coordinate quantization}

Similar to the half BPS case, we first try to quantize the
classical $R$-balls (\ref{SS}) with collective coordinate
quantization. From the analysis of the moduli space in Sec. IV, we
can parametrize this type of $R$-ball solutions in the following
way:
\begin{equation}\label{2moduli}
 Z_\alpha = \sqrt{2\theta}e^{i\varphi_\alpha}a_\alpha^V,
 \quad \alpha=1,2,
\end{equation}
where $a^V_\alpha=V^\dag a_\alpha V$ is considered as gauge
degrees of freedom labeled by $V$ in the coset space
$SU(\infty)/U(2)$ and $\varphi_\alpha$ are two independent phases.
(Here we have absorbed a factor $\exp (\pm it/R)$ into
$\varphi_{1,2}$ respectively.) Subsequently, the physical and
dynamical degrees of freedom are $\sqrt{\theta}$,
$\varphi_\alpha$, which span a reduced moduli space
$\mathbf{R}^+\times U(1)\times U(1)$ for the non-commutative $1/4$
BPS sector.

Now we measure the length in units of the radius $R$ of $S^3$; in
the large $N$ limit defined in Sec. III, we introduce $r:=
2\sqrt{c_2}$, with $c_2=N(N_1+N_2)\theta/R^2$ fixed. We also
introduce the following notions:
$${N_2\over N_1}={q\over p},$$
where $p$ and $q$ are nonnegative and coprime. Then the Lagrangian
is given by
\begin{equation}
 L={\dot{r}^2\over 2}+{r^2\over 2}(\frac{p}{p+q}\dot{\vphi}_1^2
+\frac{q}{p+q}\dot{\vphi}_1^2)-{r^2\over 2}.
\end{equation}
Gauss's law is reduced to
\begin{equation}\label{FGL}
 G:=r^2(\dot{\varphi}_1+\dot{\varphi}_2) = 0.
\end{equation}
By the standard procedure, one obtains
\begin{eqnarray}
 H&=&{p_r^2\over 2}+{1\over 2r^2}
 (\frac{p+q}{p}J_1^2 +\frac{p+q}{q}J_2^2)+ {r^2\over 2},\\
 Q&=&J_1-J_2,\\
 G&=&(p+q)\left({J_1\over p}+{J_2\over q}\right)
\end{eqnarray}
where $p_r=\dot{r}$, $J_1=pr^2\dot{\varphi}_1/(p+q)$
$J_2=qr^2\dot{\varphi}_2/(p+q)$.

Upon quantization, $J_\alpha=-i\pa/\pa\vphi_\alpha\to
m_\alpha,\;(\alpha=1,2)$, with $m_\alpha$ integers. It is easy to
see the relation of $J_\alpha$ and the Schwinger representation of
the $u(2)$ algebra: $\mathbf{L}_0=J_1+J_2$,
$\mathbf{L}_3=J_1-J_2$. Then Gauss's law dictates that
\beq{GSa}qm_1+pm_2=0\;\Rightarrow\;
m_1=pk,\;m_2=-qk,
\eeq
where $k$ is an integer. Accordingly, in the physical subspace
\begin{eqnarray}\label{H3}
 H&=&-{1\over 2r}{\partial\over\partial r}
 r{\partial\over\partial r}+\frac{(p+q)^2k^2}{2r^2}
 + {r^2\over 2}, \nn \\
 Q&=&(p+q)k.
\end{eqnarray}
The eigenstates of the Hamiltonian in (\ref{H3}) of energy $E$ are
given by
\begin{equation}\label{WFnct}
 \Psi_{nk}(r,\varphi_1,\varphi_2)=
 r^{(p+q)|k|}e^{-r^2/2}F(-n,(p+q)|k|+1,r^2)e^{ik(p\varphi_1-q\varphi_2)}
\end{equation}
where $F(.,.,r^2)$ is a confluent hypergeometric function and
$n=(E-1-(p+q)|k|)/2=0,1,2,\ldots$. Then the charge density and
energy density in units of $c_2=N(N_1+N_2)\theta/R^2$ are given by
\beq{density}
Q=(p+q)k,\hspace{0.5in} E=(p+q)|k|+2n+1.
\eeq
Ignoring the zero-point energy from the ordering ambiguity of
quantum operators, the BPS states are those corresponding to
$n=0$.

Moreover, it is obvious that the system (\ref{H3}) can be mapped
to a two-dimensional harmonic oscillator in terms of the new
variables:
\beq{2dd} \vphi=\frac{p\vphi_1-q\vphi_2}{p+q}, \hspace{0.5in}
z=re^{i\vphi}.
\eeq
Recall that $p$ and $q$ are coprime if and only if there are two
integers $s$, $t$ such that $pt-qs=1$ (B\'{e}zout's identity; it
is easy to see that $s$, $t$ are also coprime). Then the period of
$\vphi$ is actually $2\pi/(p+q)$, corresponding to
$\vphi_1\to\vphi_1+2t\pi,\;\vphi_2\to\vphi_2+2s\pi$. Consequently,
the two-dimensional harmonic oscillator is actually defined on a
cone. The spectrum of $E-Q$ from (\ref{density}) can be mapped to
that of the Landau levels labelled by $n$ with $k=0,1,2,\ldots$
labelling the degeneracy in the same Landau level (and $n$-th
``anti-Landau level'' for $k=0,-1,-2,\ldots$).

We know that an $SL(2,\mathbb{Z})$ transformation changes a pair
of coprime integers into another coprime pair; and any coprime
pair $(p,q)$ can always be generated by acting an element of
$SL(2,\mathbb{Z})$
$$\left(\begin{array}{cc} p & s \\ q & t \end{array} \right)$$
on a standard vector $(1,0)^T$. Since the coprime pair is
determined by the ratio $N_1/N_2$, the different choices for the
coprime pair correspond to different large $N$ sectors in $\CN{4}$
SYM, and $SL(2,\mathbb{Z})$ transforms between different sectors
with the same quantized $k$.

\subsection{Canonical quantization}

According to Eq.~(\ref{density}), the quantum quarter BPS state
obtained here contains essentially one quantum number $k$, hinting
that the (many-body) state is a rigid or incompressible one. This
picture is going to be checked in this subsection by applying
another quantization scheme. In contrast to the above quantization
scheme, in which the classical Gauss's law constraints except one
are solved  before quantization, now let us us start from the
classical solutions (\ref{19b}) with \ref{SS}), treat $a_\alpha
(t)$ as dynamical variables and apply canonical quantization to
them, with the non-commutativity constraints \beq{3.a.2}
[a_\alpha(t),a_\alpha^\dag(t)]= 2\theta \mathbf{1}_{N_\alpha\times
N_\alpha}, \hspace{0.6in}(\alpha=1,2), \eeq incorporated by
Langrangian multipliers. This results in the effective Lagrangian
(with two matrix Lagrangian multipliers $\lambda_\alpha$):
\beq{30} L&=&\frac{1}{2R}{\rm Tr}(D(a_1\times \mathbf{1})
D(a_1^\dag\times \mathbf{1}))+\frac{1}{2R}{\rm Tr}
(D(\mathbf{1}\times a_1) D(\mathbf{1}\times a_1^\dag)) \nn \\
&&+\frac{iN_2}{2R^2}{\rm Tr}(a_1da_1^\dag-da_1a_1^\dag)
-\frac{iN_1}{2R^2}{\rm Tr}(a_2da_2^\dag-da_2a_2^\dag) \nn \\
&&+2\theta N_2{\rm Tr}\lambda_1 -2\theta N_1{\rm Tr}\lambda_2,
\eeq where $DX=\dot{X}-i[A_0,X],\;
da_\alpha=\dot{a}_\alpha-i[\lambda_\alpha,a_\alpha]$. The direct
product structure of $Z_\alpha(t)$ ensures that the composite
operators formed by $Z_\alpha$ in SYM do not receive loop
corrections to their conformal dimensions. In the $R\to 0$ limit
we can drop the kinetic term in the Lagrangian~(\ref{30}), as long
as we focus on the ground states. Finally we end up with the
effective
matrix model: \beq{34} L&=&L_1-L_2, \nn \\
L_1&=&\frac{iN_2}{2R^2}{\rm
Tr}(a_1da_1^\dag-da_1a_1^\dag)+\frac{2N_2\theta}{R^2}{\rm
Tr}\lambda_1\nn \\ L_2&=&\frac{iN_1}{2R^2}{\rm
Tr}(da_2a_2^\dag-a_2da_2^\dag)+\frac{2N_1{\theta}}{R^2}{\rm
Tr}\lambda_2. \eeq We see that $L_1$ or $L_2$ is separately a
NCCSMM model that has been discussed by Susskind
\cite{Susskind01a}:
\beq{35} L=\frac{i\xi}{2R^2}{\rm Tr}(udu^\dag-duu^\dag)
+\frac{2\xi{\theta}}{R^2}{\rm Tr}\lambda,
\eeq
where $\xi=N_2$ or $N_1$, depending on whether $u=a_1$ or
$u=a_2^\dag$. This NCCSMM model has a different origin, compared
with the matrix model in ref. \cite{Yamada05}, where a chemical
potential $\mu$ was introduced in $\CN{4}$ SYM as an external
parameter.

To quantize, we introduce the operators ${\bf x}_{ij}$ and ${\bf
y}_{ij}$ through  ${\bf u}_{ij}={\bf x}_{ij}+i{\bf y}_{ij}$, and
impose the canonical commutation relations: \beq{36}[{\bf
x}_{ij},{\bf y}_{mn}]= i\frac{R^2}{N\xi}\delta_{in}\delta_{jm}.
\eeq Here we used the fact that $N^{-1}$ plays the role of the
Planck constant in a large $N$ matrix model. The non-commutative
constraints are now imposed on the physical states:
\beq{37} ({\bf
x}_{ij}{\bf p}_{jm}-{\bf p}_{ij} {\bf x}_{jm}) |\Psi\rangle
=\frac{i}{\nu}\delta_{im}|\Psi\rangle,
\eeq
where $\nu=R^2/N\xi{\theta}$ and ${\bf p}_{ij}={\xi\bf
y}_{ij}/R^2$.

The left side of (\ref{37}) resembles the angular momentum
operator in quantum mechanics, which generates a rotation of
particles in a two-dimensional plane. We may define a unitary
matrix to generate such a rotation:
\beq{38} \hat{T}=\exp\{i\omega_{im} ({\bf
x}_{ij}{\bf p}_{jm} -{\bf p}_{ij}{\bf x}_{jm})\},
\eeq
with $\omega_{im}$ the angles of rotation. As we consider the
operation to exchange two particles, i.e., to rotate them by an
angle ${\rm Tr}\omega=\pi$, we have
\beq{39}
\hat{T}|\Psi\rangle=e^{i\pi/\nu}|\Psi\rangle.
\eeq
This indicates that the many-body state $|\Psi\rangle$ is a QH
state of fermions when $1/\nu$ is odd, or of bosons when $1/\nu$
is even, where $1/\nu$ is just the filling fraction of the QH
system. The well-known quantization of $\nu$ in the NCCSMM model
implies $\theta=k/N\xi R^2$, with $k$ a positive integer.

Applying the above results to our matrix model~(\ref{34}), we have
\beq{41} \left. {\nu_1^{-1}=k_1=NN_2\theta/R^2, \atop
\nu_2^{-1}=k_2=NN_1\theta/R^2,}\right\}\quad \Rightarrow \quad
k_1=qk,\;k_2=pk, \eeq where $k$ is an integer, and $(p,q)$ is
again a pair of coprimes defined by the ratio $N_2/N_1=q/p$.
Substituting the quantized $\theta$ into Eq.~(\ref{ES}), we obtain
the quantized $R$-charge \beq{QR} Q=c_2=(p+q)k. \eeq It is the
same as we obtained before from collective coordinate
quantization.

\subsection{New Higher Dimensional Quantum Hall State}

In the above we have constructed mathematically a new quantum BPS
state in $\CN{4}$ SYM, with energy equal to its $R$-charge. What
is the physical interpretation of this state? It is known that the
NCCSMM model (\ref{35}) describes a FQH system with filling factor
$\nu=1/k$. So the ground state of our model (\ref{34}) describes a
quantum state that is the product of the two FQH states,
respectively, on two orthogonal non-commutative planes. To get
some ideas about what it looks like, we note that for the
classical $R$-ball solution~(\ref{SS}), the constant matrices
$A_1$ and $A_2$ are neither Hermitian nor normal. However, one may
form two Hermitian matrices from them: $Z_1Z_1^\dag=A_1A_1^\dag$
and $Z_2Z_2^\dag=A_2A_2^\dag$, which can be diagonalized
simultaneously by a unitary rotation, since one can easily verify
$[A_1A_1^\dag,A_2A_2^\dag]=0$. It is not hard to see that there
are only $N_1$ independent eigenvalues of $A_1A_1^\dag$,
$|a_{1,i}|^2,\;(i=1,2,...,N_1)$, and $N_2$ independent eigenvalues
of $A_2A_2^\dag$, $|a_{2,j}|^2,\;(j=1,2,...,N_2)$. Upon
quantization, $|a_{1,i}|^2$ and $|a_{2,j}|^2$ can be interpreted,
respectively, as the radial positions (squared) of $G$-particles
in $Z_1$- and $Z_2$-plane. Here as in the commutative half BPS
case, we adopt the interpretation of matrix diagonal elements as
coordinates of particles, which we have named as $G$-particles, in
the same spirit as the BFSS matrix model \cite{BFSS}. Note that
$Z_1$ and $Z_2$ are $N$-by-$N$ matrices with $N=N_1N_2$, which is
just the number of $G$-particles. So the distribution of $N$
$G$-particles in four-dimensional internal space thus forms a
rectangular lattice with spacing $\theta$ in the two-dimensional
plane spanned by $|a_1|^2$ and $|a_2|^2$ (Fig.~1). Clearly there
are $N_1$ columns of $G$-particles (Fig.~1) distributing along
$|a_1|^2$ direction, and each of them has $N_2$ rows of
$G$-particles (Fig. 1) along $|a_2|^2$-direction. Altogether there
are $N_1N_2$ $G$-particles distributed on a four dimensional
space, that is the product of two circular disks, respectively, on
non-commutative $Z_1$- and $Z_2$-plane. Each column (or row)
represents a quantum Hall droplet. So their direct product
represents a higher dimensional quantum Hall state in four
dimensions. (It is shown that the Landau Hamiltonian in flat space
with even dimensions can be reduced to the direct sum of two
dimensional Landau Hamiltonians \cite{Meng03}.)
\begin{figure}[hpt]
\setlength{\unitlength}{1cm}
\begin{picture}(6,6)
\put(0,1){\vector(1,0){6}} \put(1,0){\vector(0,1){5}}
\multiput(1,1)(0.3,0){15}{\circle*{0.1}}
\multiput(1,1.3)(0,0.3){9}{\begin{picture}(0,0)
 \multiput(0,0)(0.3,0){15}{\circle*{0.1}}
  \end{picture}}
\put(5.7,0.6){\mbox{$|a_1|^2$}} \put(0.1,4.8){\mbox{$|a_2|^2$}}
\end{picture}
\begin{minipage}{5in}
\caption{The distribution of $N$ $G$-particles in ($|a_1|^2$,
$|a_2|^2$) space. The lattice spacing is $\theta$.}
\end{minipage}
\end{figure}
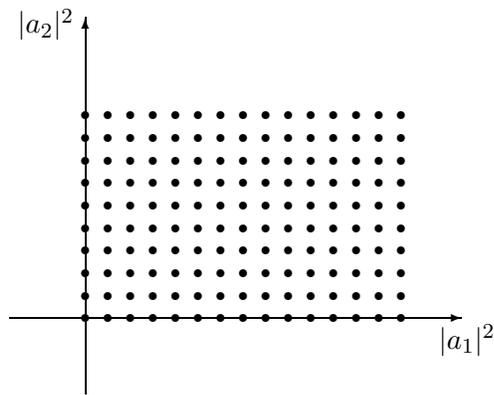

What would be the holographic correspondence of the new $1/4$ BPS
states given by Eq.~(\ref{density}) or (\ref{QR}) in the IIB
string dual? The above figure motivates us to suggest the
following picture: The $G$-particles, thought of as ``sources'' in
IIB supergravity like LLM particles in the LLM construction, form
a four dimensional object: namely we have a bunch ($N_2$) of FQH
droplets, each living on the $Z_1$-plane, consisting of $N_1$
$G$-particles and looking like a point-like object on the
$Z_2$-plane; and the bunch of $N_2$ point-like objects also form a
FQH droplet on the $Z_2$-plane. This state is certainly not the
FQH state of LLM fermions that the present authors suggested in
the IIB quantum gravity about a year ago\cite{DWW05}, which is
known as a deformation of the $1/2$ BPS IIB geometry to have null
singularity. The above picture for the states (\ref{density})
immediately suggests themselves as a resolution of our previously
proposed FQH states of LLM fermions: Namely in the limit $N_1,N_2
\to \infty$ with fixed $N_2/N_1=q/p\ll 0$, the quantum states
(\ref{density}) become a candidate for the SYM dual of the FQH
states in IIB gravity suggested by us \cite{DWW05}. Indeed, we can
present several evidences for this suggestion:

\begin{itemize}
\item We may impose an extra condition, $g^2N_2\sim{\rm fixed}$,
According to the standard AdS/CFT dictionary, the typical length
scale along $Z_1$-plane is $l_s(g^2N_1)^{1/4}\gg l_s$ with $l_s$
the stringy scale, while the typical scale along $Z_2$ plane is
$l_s(g^2N_2)^{1/4}\sim l_s$. Therefore, the classical geometry
along $Z_1$-plane is well-defined, and we can identify this plane
as the boundary plane of LLM geometries. Meanwhile, the classical
geometry description along $Z_2$-plane breaks down, and the
quantum corrections play a role at the string scale and resolve
the original singularity.
\item The angular momentum or $R$-charge contributed by $Z_1$
and $Z_2$ are proportional to $pkN_1^2$ and $pkN_1N_2$,
respectively. The area occupied by a FQH droplet is proportional
to its angular momentum. Hence the typical size of the quantum
states is $N_1R$ along $Z_1$-plane and $\sqrt{N_1N_2}R$ along
$Z_2$-plane. If we take $N_1R\sim$~fixed as a macroscopic scale,
$\sqrt{N_1N_2}R\to 0$ will be a microscopic scale. Then the
configuration~(\ref{19b}) looks like a thin pancake in internal
four-dimensional space (Fig.~2), and the states looks like a
two-dimensional incompressible fluid (FQH fluid) macroscopically.
The thin thickness of such FQH fluid in the transverse directions
can be understood as a necessity for the resolution of null
singularity in dual string theory.

\begin{figure}[hptb]
\centering
\includegraphics[width=3in]{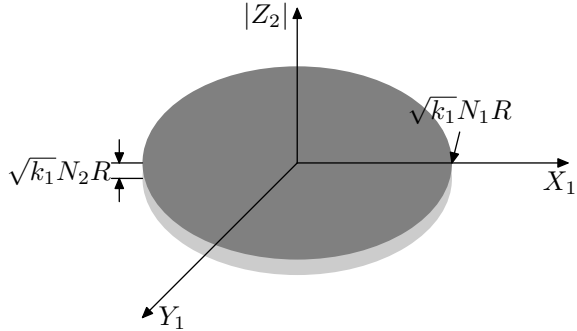}
\begin{minipage}{4in}
 \caption{The semi-classical configuration of a four dimensional
 fractional quantum Hall state. $|Z_2|$ denotes the radius
 in the transverse 2-plane.}
\end{minipage}
\end{figure}

\item The Hamiltonian $H$ or $R$-charge is a simple summation,
$H=H_1+H_2$, where $H_1$ and $H_2$ denoted the contribution from
$Z_1$ and $Z_2$ respectively. Since $Q_2/Q_1=q/p\ll 0$, $H_2$ can
be treated as a perturbation. At the zeroth order, we have only
$H_1$, which describes a 2d FQH system, as what we have suggested
in ref. \cite{DWW05}.
\end{itemize}

Of course, our result showed that this resolution actually breaks
more supersymmetries. Because $\CN{4}$ SYM is a well-defined
quantum theory that is believed to contain complete information on
IIB superstring theory, our study suggests a possible way to deal
with the properties of the spacetime geometries near null
singularities, or emergent geometry \cite{Ber05b}, in terms of
dual quantum field theory or its reductions to matrix models.

To conclude this subsection, we make two remarks: First, it would
be interesting to see whether a smooth IIB geometry could be
generated by such ``seeds'' in the IIB supergravity dual. We note
that the solution (\ref{19b}) and the above quantization procedure
preserve the isometry group $SO(2)\times SO(2)\times R$. Second,
the statistics we were talking about in the last subsection is the
statistics of $N_1$ $G$-particles in one fixed FQH droplet on
$Z_1$-plane and the statistics of the $N_2$ FQH droplets as
identical objects when two of them are exchanged on $Z_2$-plane.
Because of the direct product structure of the total quantum
state, interchanging any two of $N_1N_2$ $G$-particles is not an
admissible symmetry operation.

\section{Summary and Discussions}

A central issue for understanding AdS/CFT holography is to see how
geometry or gravity emerges in the CFT dual. In particular, one
wants very much to see how the LLM fermions in a quantum Hall
droplet, that are known \cite{LLM} to ``encode'' a wide class of
half BPS IIB geometries, arise in the dual gauge theory. In this
paper we have proposed a new framework for constructing quantum
candidate states in $\CN{4}$ SYM on $R\times S^3$, which are
promising for holographically encoding classical or quantum
geometries on the gravity side.

In our proposal, these candidates are {\em quantized} $R$-ball
states, with energy saturated by a conserved $U(1)$ $R$-charge.
They are constructed by quantization over the moduli space of
certain classical $R$-balls, which are spatially constant,
time-dependent (rotating in internal space) and maintain a
fraction of supersymmetries.  Many features of the Berenstein's
matrix model for the commutative half BPS sector emerge naturally
in our framework with space-filling $R$-balls. In particular, the
origin of the ``magnetic field'' in the QH analogy is identified
to be the rotation of the $R$-balls in internal space that
generates $R$-charge, and the origin of the projection down to the
lowest Landau level is closely related to the BPS bound (the
energy is saturated by the $R$-charge). Quantization of such
$R$-balls results in a many-body quantum Hall system with filling
factor $\nu=1$, whose constituents can be identified with the LLM
fermions. The system is a non-interacting one, whose constituents
are called $G$-particles with their number related to the rank $N$
of the color gauge group. Gauge invariance (the Gauss's law) plays
an important role in reducing the number of physical degrees of
freedom. In the half BPS case, this reduces the degrees of freedom
in the physical quantum states from $N^2$ to $N$.

The success in making the QH analogy of the half BPS dynamics
meaningful and substantial encouraged us to look for FQH-like
states in the quarter BPS sector. In our framework, we have been
able to shown that non-commutative almost BPS classical $R$-balls
are allowed in the large $N$ limit, with two non-vanishing complex
scalars. Upon quantization they lead to a NCCSMM model that
describe the ``direct product'' of two QH droplets on a pair of
orthogonal planes, each in $\nu=1/k$ FQH states ($k$ being an
integer). Thus the quantum states are those of an interacting
many-body system, actually a new four-dimensional QH system. In a
special limit, the states reduces approximately to the FQH states
on a plane that correspond to the incompressible giant graviton
fluid (with density $\rho=1/k$ proposed by the present authors
\cite{DWW05} previously in the dual gravity theory). The latter
was known to give rise to geometry with null singularities, and we
interpret the four dimensional new QH states obtained as
representing a resolution of the null singularities in the quantum
theory of gravity.  (Note that the $d=4$ QH effect we have here is
{\em not} as the $d=4$ QHE proposed previously in ref.
\cite{HuZhang,KN}, which are not a direct product of two Abelian
QHE.)

We note that both the compactness of the space $S^3$ and
non-commutativity permitted in the large $N$ limit $N\to\infty$
play an essential role in admitting the existence of the new
(FQH-like) $R$-ball states. First, the conformal coupling term,
that couples the scalars to the spatial curvature, gives rise to a
harmonic confining potential over the usual moduli space of vacua
in Minkowski spacetime. Second, $S^3$ has a finite volume.
Combining these two facts, it makes sense to consider
space-filling $R$-balls with a rotation in internal space
generating an $R$-charge and to examine the small radius $R\to 0$
limit. This limit allows us not only to single out the lowest
Kaluza-Klein modes, but also to project the $R$-balls down to LLL.
This is because with a particular rotation frequency, the
centrifugal force just cancels the harmonic confining potential,
leading to the LLL (or BPS) states. Finally, the non-commutative
$R$-balls we found before that upon quantization exhibit FQH-like
behavior exist and become BPS only in the large $N$ limit, which
is just the defining limit for AdS/CFT holography. This suggests
to us that non-commutative geometry should play a profound role in
studying the emergent gravity in the holographical CFT dual.

On the physics side, conceptually our study has heavily explored
the analogy with two recent inter-related developments (BEC and
QHE) in many-body systems. This is not surprising, since string/M
theory essentially is a many-body system from the point of view of
the BFSS matrix model \cite{BFSS}. The present authors hold the
belief that the string/M theory has a profound connection with
strongly correlated systems that are one of the recent focuses of
attention in quantum many-body physics. One important concept is
that of BEC, which plays a crucial role in the present context:
The $R$-balls can be viewed as a rotating BEC on $S^3$ and what we
have studied is the dynamics of a rapidly rotating BEC. Related to
the rotating BEC is the QH effect for fermions, both integral and
fractional, at small boson filling fractions \cite{Wilkin01}.
Previously a possible realization of the QHE in string theory has
been proposed in the literature \cite{Susskind2}, which involved
particular configurations of certain branes. Our present study
suggests a more fundamental and ubiquitous connection of string
theory with the QHE in particular and with non-commutative
geometry in general.

\acknowledgments{J. Dai thanks V. P. Nair, A. Polychronakos, D.
Karabali and A. Agarwal, and X.-J. Wang thanks Yi-Hong Gao,
Jian-Xin Lu and Miao Li, for helpful discussions. J. Dai was
partly supported by a CUNY Collaborative Research Incentive grant.
X.-J. Wang was partly supported by China NSF, Grant No. 10305017,
and through USTC ICTS by grants from the Chinese Academy of
Science and a grant from NSFC of China. Y.-S. Wu was supported in
part by the US NSF through grant PHY-0457018.}

\appendix

\section{$\CN{4}$ supersymmetry algebra on $R\times S^3$}

By a straightforward calculation, the supersymmetry
transformation~(\ref{susy}) leads to the following supercurrents:
\beq{22} \bar{J}^\mu_L&=&-\frac{i}{2}{\rm
Tr}(\bar{\psi}F_{\rho\sigma}) \gamma^\mu \gamma^{\rho\sigma}-{\rm
Tr}\{\bar{\psi} (\alpha^iD_\nu
X_i+i\gamma_5\beta^jD_\nu Y_j)\}\gamma^\mu\gamma^\nu \nn \\
&&+\frac{1}{2}\ep^{ijk}{\rm Tr}\{\bar{\psi}
(\alpha^k[X_i,X_j]+\beta^k[Y_i,Y_j])\}\gamma^\mu -i{\rm
Tr}(\bar{\psi} [X_i,Y_j])\alpha^i\beta^j\gamma^\mu\gamma_5 \nn \\
&&-\frac{i}{R}{\rm
Tr}\{\bar{\psi}(X_i+i\gamma_5\beta^jY_j)\}\gamma^\mu
\gamma_5\gamma^0, \nn \\
\bar{J}^\mu_R&=&\bar{J}^\mu_L+\frac{2i}{R}{\rm
Tr}\{\bar{\psi}(X_i+i\gamma_5\beta^jY_j)\}\gamma^\mu
\gamma_5\gamma^0. \eeq In a curved space with constant curvature,
global supercharges associated to the above supercurrents can be
defined by appropriately projecting the locally-defined
$\bar{J}^0$ to a global section. To this end, we introduce the
transformation
\beq{22a} \zeta_{L}=M_{L}\zeta_{L0}, \hspace{0.5in}
\zeta_{R}=M_{R}\zeta_{R0}, \eeq where the $4\times 4$ matrices
$M_{L,R}$ depend on spacetime coordinates and $\zeta_{L0}$ and
$\zeta_{R0}$ are two constant Majorana spinors. Then the global
supercharges can defined with the help of $\zeta_{L0}$ and
$\zeta_{R0}$: \beq{23}
\bar{Q}_L=\int_{S^3}\bar{J}^0_LM_L,\hspace{0.6in}
\bar{Q}_R=\int_{S^3}\bar{J}^0_RM_R. \eeq

We take the metric of $R \times S^3$ to be
$$ds_2^2=dt^2-d\theta^2-\sin^2\theta (d\psi^2+\sin^2\psi d\chi^2).$$
Accordingly, the explicit solution of the conformal Killing spinor
equation~(\ref{5}) is given by \cite{LPR98,Okuyama02}, \beq{24}
\ep=e^{\frac{it}{2R}\Gamma_0}
e^{\frac{i\theta}{2R}\Gamma_{15}}e^{-\frac{\psi}{2R}\Gamma_{12}}
e^{-\frac{\chi}{2R}\Gamma_{23}}\ep_0, \eeq where $\ep_0$ is a
constant spinor, and $\Gamma^a$ denotes $\gamma$-matrices in the
local Lorentzian frame of $AdS_5$. Eq.~(\ref{24}) together with
Eq.~(\ref{6}) lead to \beq{25}
M_L&=&e^{it/R}e^{-\frac{i\theta}{2R}\gamma_{01}}
e^{-\frac{\psi}{2R}\gamma_{12}}e^{-\frac{\chi}{2R}\gamma_{23}},
\nn \\
M_R&=&e^{-it/R}e^{\frac{i\theta}{2R}\gamma_{01}}
e^{-\frac{\psi}{2R}\gamma_{12}}e^{-\frac{\chi}{2R}\gamma_{23}}.
\eeq Here $\gamma_{ab}$, as we noted in Section II, are defined in
the local Lorentzian frame on $R \times S^3$.

We will focus on the fermionic part of the superconformal algebra
that involves only the charges of $R$-ball configurations. In the
$A_0=0$ gauge the variation of the supercurrent can be written as
follows: \beq{26}
\delta_L\bar{J}^0_L&=&-2iT^{0\nu}\bar{\zeta}_L\gamma_\nu
-\frac{2i}{R}\bar{\zeta}_L\gamma^0\alpha^i\beta^j{\rm Tr}
(X_i\dot{Y}_j-\dot{X}_iY_j) \nn \\
&&+\frac{2}{R}\ep^{ijk}\bar{\zeta}_L\gamma^0\gamma_5(\alpha^k{\rm
Tr}X_i\dot{X}_j+\beta^k{\rm Tr}Y_i\dot{Y}_j)+..., \nn \\
\delta_R\bar{J}^0_R&=&-2iT^{0\nu}\bar{\zeta}_R\gamma_\nu
+\frac{2i}{R}\bar{\zeta}_R\gamma^0\alpha^i\beta^j{\rm Tr}
(X_i\dot{Y}_j-\dot{X}_iY_j) \nn \\
&&-\frac{2}{R}\ep^{ijk}\bar{\zeta}_R\gamma^0\gamma_5(\alpha^k{\rm
Tr}X_i\dot{X}_j+\beta^k{\rm Tr}Y_i\dot{Y}_j)+..., \\
\delta_L\bar{J}^0_R&=&-2iT^{0\nu}\bar{\zeta}_L\gamma_\nu+\frac{2i}{R^2}
\bar{\zeta}_L\gamma^0{\rm Tr}(X_i^2+Y_j^2)-
\frac{2}{R}\bar{\zeta}_L\gamma^0\gamma_5{\rm Tr}(X_i\dot{X}_i+
Y_j\dot{Y}_j)+..., \nn \\
\delta_R\bar{J}^0_L&=&-2iT^{0\nu}\bar{\zeta}_R\gamma_\nu+\frac{2i}{R^2}
\bar{\zeta}_R\gamma^0{\rm Tr}(X_i^2+Y_j^2)+
\frac{2}{R}\bar{\zeta}_R\gamma^0\gamma_5{\rm Tr}(X_i\dot{X}_i+
Y_j\dot{Y}_j)+\cdots  \ . \nn \eeq Here $T^{\mu\nu}$ is the
energy-momentum tensor obtained from the SYM Lagrangian~(\ref{1}).
The $\cdots$ terms may involve scalars of higher degrees, such as
\beq{26a} \ep^{ijk}\bar{\zeta}\gamma_5\{\alpha^k\beta^l {\rm
Tr}([X_i,Y_l]X_j) +i\gamma_5\alpha^l\beta^k {\rm
Tr}([X_l,Y_i]Y_j)\}. \eeq We have checked that for the $R$-balls
we obtained in the text, whether commutative or not, the terms
presented in (\ref{26}) are the only non-vanishing ones. For
example, it is easy to verify that the terms in (\ref{26a})
vanishes for both $1/2$ and $1/4$ BPS $R$-ball solutions obtained
in the text.

We introduce the following ``matrix charges'': \beq{27}
P_{L,R}&=&\int_{\mathbf{S}^3}\gamma^0M_{L,R}^\dag\gamma^0\gamma_\nu
M_{L,R}T^{0\nu}
=P_{L,R}^a\Gamma_a+S_{L,R}^a\Gamma_a\Gamma_5, \nn \\
K&=&\int_{\mathbf{S}^3}\gamma^0M_{L}^\dag\gamma^0\left[\gamma_\nu
T^{0\nu} -\frac{1}{R^2}\gamma^0{\rm Tr}(X_i^2+Y_j^2)\right]M_{R},
\eeq and the ``angular momenta'': \beq{27a}
L_{ij}&=&\frac{1}{R}\int_{\mathbf{S}^3}M_{L,R}^\dag M_{L,R} {\rm
Tr} (X_i\dot{Y}_j-\dot{X}_iY_j)=\frac{1}{R}\int_{\mathbf{S}^3}
{\rm Tr} (X_i\dot{Y}_j-\dot{X}_iY_j),  \nn \\
L_X^k&=&\frac{1}{R}\ep^{ijk}
\int_{\mathbf{S}^3}\gamma_5M_{L,R}^\dag\gamma_5M_{L,R} {\rm
Tr}X_i\dot{X}_j=\frac{1}{R}\ep^{ijk} \int_{\mathbf{S}^3}
{\rm Tr}X_i\dot{X}_j, \nn \\
L_Y^k&=&\frac{1}{R}\ep^{ijk}
\int_{\mathbf{S}^3}\gamma_5M_{L,R}^\dag\gamma_5M_{L,R} {\rm
Tr}Y_i\dot{Y}_j =\frac{1}{R}\ep^{ijk} \int_{\mathbf{S}^3} {\rm
Tr}Y_i\dot{Y}_j. \eeq Then with $\delta {\cal O}
=-i\{\bar{\zeta}_0Q,O\}$, the superconformal algebra can be
written symbolically as \beq{28} \{Q_L,\bar{Q}_L\}&=&2P_L
+2L_{ij}\alpha^i\beta^j\gamma^0
+2i\gamma^0\gamma_5(\alpha^kL_X^k+\beta^kL_Y^k)+\cdots, \nn \\
\{Q_R,\bar{Q}_R\}&=&2P_R -2L_{ij}\alpha^i\beta^j\gamma^0
-2i\gamma^0\gamma_5(\alpha^kL_X^k+\beta^kL_Y^k)+\cdots, \nn \\
\{Q_L,\bar{Q}_R\}&=&2K+\cdots,\hspace{1in}
\{Q_R,\bar{Q}_L\}=2K^\dag+\cdots. \eeq Here we have presented only
terms that are relevant in this paper. The emergence of $\gamma^0$
in the terms involving ``angular momenta'' indicates that $L$'s
are not central charges. Because matrices $\alpha,\;\beta$
generate a rotation among indices of $SU(4)$ R-symmetry group,
these $L$'s are actually associated with the R-charges. We may
introduce fermionic charges $Q$ and $S$ by $Q_L=Q+S,\;Q_R=Q-S$.
Then schematically the algebra (\ref{28}) can be rewritten as
\beq{28a} \{Q,\bar{Q}\}&=&P_L+P_R+(K+K^\dag)+\cdots,\hspace{0.6in}
\{S,\bar{S}\}=P_L+P_R-(K+K^\dag)+\cdots, \nn \\
\{Q,\bar{S}\}&=&(K-K^\dag)+2L+\cdots,\hspace{1in}
\{S,\bar{Q}\}=-(K-K^\dag)+2L+\cdots. \eeq As the radius $R$ of
$S^3$ goes to infinity, we have $P_{L,R},K,K^\dag\to \sla{p}$ with
$p$ the four momentum in four-dimensional Minkowski spacetime,
$M^4$. Therefore, the superconformal algebra~(\ref{28a}) reduces
to the standard form in $M^4$.

The general expression for the BPS bound can be obtained by
computing the eigenvalues of the right-hand side of the algebra
(\ref{28}), but the computation would be very tedious due to the
presence of many off-diagonal elements. Here we only focus on the
BPS bound for the supersymmetric configurations found in the text.
It is not hard to see that we always have $K=0$ for these
backgrounds. For the commutative half BPS $R$-balls, with only one
pair of scalar and pseudo-scalar turned on, we have $L_X=L_Y=0$.
Meanwhile, for our non-commutative $R$-balls, $L_X,\;L_Y\propto
{\rm Tr}(a_1\times a_2)=0$ by using (\ref{32}). Notice that
$T^{0i}=0$ for these configurations. Computing the eigenvalues on
the right side of the superconformal algebra (\ref{28}), we obtain
the BPS bound for energy: \beq{29} E\geq |L|, \eeq where the
angular momentum $L$ is nothing but just $R$-charge $Q_{\mathbf
r}$ defined in eq, (\ref{BE}). For commutative half BPS $R$-balls,
this BPS bound is exactly saturated, while for the non-commutative
$1/4$ BPS $R$-balls, their energy receives an extra contribution
from the tree-level quartic interactions, which is of order
$\lambda/N(N_1+N_2)$ and can be ignored in the limit $N\to\infty$.

\end{document}